\documentclass[runningheads]{llncs}

\usepackage{array}
\usepackage{algorithm}
\usepackage{algpseudocode}
\usepackage{amsmath}
\usepackage{amsfonts}
\usepackage{amssymb}
\usepackage{graphicx}
\usepackage{caption}
\usepackage{epstopdf,epsfig}
\usepackage{xcolor}
\usepackage{balance}
\usepackage{colortbl}
\usepackage{url}
\usepackage{algpseudocode}
\usepackage{booktabs}
\usepackage{algorithmicx}
\usepackage{pbox}
\usepackage{float}
\usepackage{dblfloatfix}
\usepackage{enumitem}
\usepackage{bm}
\usepackage[font=scriptsize,labelfont={bf,sf}]{subfig}
\newcolumntype{L}[1]{>{\raggedright\let\newline\\\arraybackslash\hspace{0pt}}m{#1}}
\newcolumntype{C}[1]{>{\centering\let\newline\\\arraybackslash\hspace{0pt}}m{#1}}
\newcolumntype{R}[1]{>{\raggedleft\let\newline\\\arraybackslash\hspace{0pt}}m{#1}}

\begin{document}

\title{Analyzing Defense Strategies Against Mobile Information Leakages: A Game-Theoretic Approach}
\author{Kavita Kumari\inst{1}, Murtuza Jadliwala\inst{1}, \\Anindya Maiti\inst{1}, and Mohammad Hossein Manshaei\inst{2}}
\authorrunning{Kumari et al.}
\institute{{$^1$University of Texas at San Antonio}\\
{$^2$Isfahan University of Technology}\\
\email{kavita.kumari@utsa.edu, murtuza.jadliwala@utsa.edu, \\a.maiti@ieee.org, manshaei@cc.iut.ac.ir}}

\maketitle

\begin{abstract}
Abuse of zero-permission sensors (e.g., accelerometers and gyroscopes) on-board mobile and wearable devices to infer users' personal context and information is a well-known privacy threat, and has received significant attention in the literature. At the same time, efforts towards relevant protection mechanisms have been ad-hoc and have main focused on threat-specific approaches that are not very practical, thus garnering limited adoption within popular mobile operating systems.
It is clear that privacy threats that take advantage of unrestricted access to these sensors can be prevented if they are effectively regulated. However, the importance of these sensors to all applications operating on the mobile platform, including the dynamic sensor usage and requirements of these applications, makes designing effective access control/regulation mechanisms difficult. 
Moreover, this problem is different from classical intrusion detection as these sensors have no system- or user-defined policies that define their authorized or correct usage.  
Thus, to design effective defense mechanisms against such privacy threats, a clean slate approach that formalizes the problem of sensor access (to zero-permission sensors) on mobile devices is first needed.
The paper accomplishes this by employing game theory, specifically, signaling games, to formally model the strategic interactions between mobile applications attempting to access zero-permission sensors and an on-board defense mechanism attempting to regulate this access.
Within the confines of such a formal game model, the paper then outlines conditions under which equilibria can be achieved between these entities on a mobile device (i.e., applications and defense mechanism) with conflicting goals.
The game model is further analyzed using numerical simulations, and also extended in the form of a repeated signaling game.

\end{abstract}

\section{Introduction}
\label{sec:Intro}

Modern mobile and wearable devices, equipped with state-of-the-art sensing and communication capabilities, enable a variety of novel context-based applications such as social networking, activity tracking, wellness monitoring and home automation. The presence of a diverse set of on-board sensors, however, also provide an additional attack surface to applications intending to infer personal user information in an unauthorized fashion. In order to thwart such privacy threats, most modern mobile operating systems (including, Android and iOS) have introduced stringent access controls on front-end or user-accessible sensors, such as microphone, camera and GPS. As a result, the focus of adversarial applications has now shifted to employing on-board sensors that are not guarded by strong user or system-defined access control policies. Examples of such back-end or user-inaccessible sensors include accelerometer, gyroscope, power meter and ambient light sensor, and we refer to these as \emph{zero-permission sensors}. As all installed applications have access to them by default, and that they cannot be actively disengaged by users on an application-specific basis, these zero-permission sensors pose a significant privacy threat to mobile device users, as has been extensively studied in the security literature \cite{FeltFCHW:2011,SchlegelZZIKW:2011,CaiC:2011,OwusuHDPZ:2012,MarquardtVCT:2012,NguyenCWBZ:2012,HanONPZ:2012,MiluzzoVBC:2012,GaoFSKYL:2014,MichalevskyDG:2014,michalevsky2015powerspy,narain2016inferring,Liu:2015:GBE:2810103.2813668,wang2015,maiti:2016,Wang:2016:FFY:2897845.2897847,MaitiJHB:2018,SabraMJ:2018,MaitiHSJ:2018}. 


At the same time, efficient and effective protection mechanisms against such privacy threats is still an open problem \cite{cai2009defending}. One of the main reasons why zero-permission sensors have limited or no access control policies associated with them is because they are required by all applications (accessed by means of a common set of libraries or APIs) primarily for efficient and user-friendly operation on the device's small and constrained form factor and display. 
For instance, gyroscope data is used by applications to re-position front-ends (or GUIs) depending device orientation, while an ambient light sensor is used to update on-screen brightness. Thus, a straightforward approach of completely blocking access or reducing the frequency at which applications can sample data from these sensors is not feasible, as it will significantly impact their usability. Alternatively, having a static access control policy for each application is also not practical as it will become increasingly complex for users to manage these policies. Moreover, such an approach will not protect against applications that gain legitimate access to these sensors (based on such static policies). Given that all applications (with malicious intentions or not) can request access to these sensors without violating any system security policy, an important challenge for a defense mechanism is to differentiate between authentic sensor access requests and requests that could be potentially misused. 

In order to begin addressing this long-standing open problem, we take a clean-slate approach by first formally (albeit, realistically) modeling the strategic interactions between (honest or potentially malicious) mobile applications and an on-board defense mechanism that cannot differentiate between their (sensor access) requests. 
We employ \emph{game-theory} as a vehicle for modeling and analyzing these interactions. Specifically, we model the following scenario. A defense mechanism on a mobile operating system receives requests to access zero-permission sensors from two different \emph{types} of applications: \emph{honest} and \emph{malicious}. Each of these applications could send either a \emph{normal} or a \emph{suspicious} request for access to on-board zero-permission sensors. A request could be classified as suspicious or normal (non-suspicious) based on the context, frequency or amount of requested sensor data. Although honest applications would typically make normal requests, they could also make suspicious requests depending on application- or context-specific operations and requirements. They could also make suspicious requests to improve overall application performance and usability. The goal of malicious applications, on the other hand, is to successfully infer private user data from these requests. Normal requests would give them some (probably, not enough) data to carry out these privacy threats, however, suspicious requests could give them additional critical data either to amplify or increase the success probability of their attacks. The defense mechanism, on receiving the request, has one of the following two potential responses: (i) \emph{accept} the request and release the requested sensor data, or (ii) \emph{block} the request preventing any data being released to the requesting application. 
It should be noted that the defense mechanism does not know the type of the application (i.e., honest or malicious) sending a particular request (i.e., suspicious or non-suspicious), as all mobile applications can currently request zero-permission sensor data without raising a flag or violating any policy.
In other words, the defense mechanism has \emph{imperfect information} on the type of application sending the request. The requesting application, on the other hand, has perfect information about its type and potential strategies of the defense mechanism.  
Given this scenario, the following are the main technical contributions of this paper:
\begin{enumerate}
\item We first formally model the strategic interactions between mobile applications and a defense mechanism (outlined above) using a \emph{two-player}, \emph{imperfect-information} game, called the \emph{signaling game} \cite{cho1987signaling}. We refer to it as the \emph{Sensor Access Signaling Game}.
\item Next, we solve the Sensor Access Signaling Game by deriving both the pure- and mixed-strategy \emph{Perfect Bayesian Nash Equilibria (PBNE)} strategy profiles possible in the game.
\item Finally, by means of numerical simulations, we examine how the obtained game solutions or equilibria evolve with respect to different system (or game) parameters in both the \emph{single-stage} and \emph{repeated} (more practical) scenarios.
\end{enumerate}
Our game-theoretic model, and the related preliminary results, is the first clean-slate attempt to formally model the problem of protecting zero-permission sensors on mobile platforms against privacy threats from strategic applications and adversaries (with unrestricted access to it). Our hope is that this model will act as a good starting point for designing efficient, effective and incentive-compatible strategies for protecting against such threats.




\section{Sensor Access Signaling Game}
\label{sec:sysmodel}

\begin{figure}[t]
\centering
\subfloat[System Model.\label{fig:sysmodel}]{
\includegraphics[width=.50\textwidth]{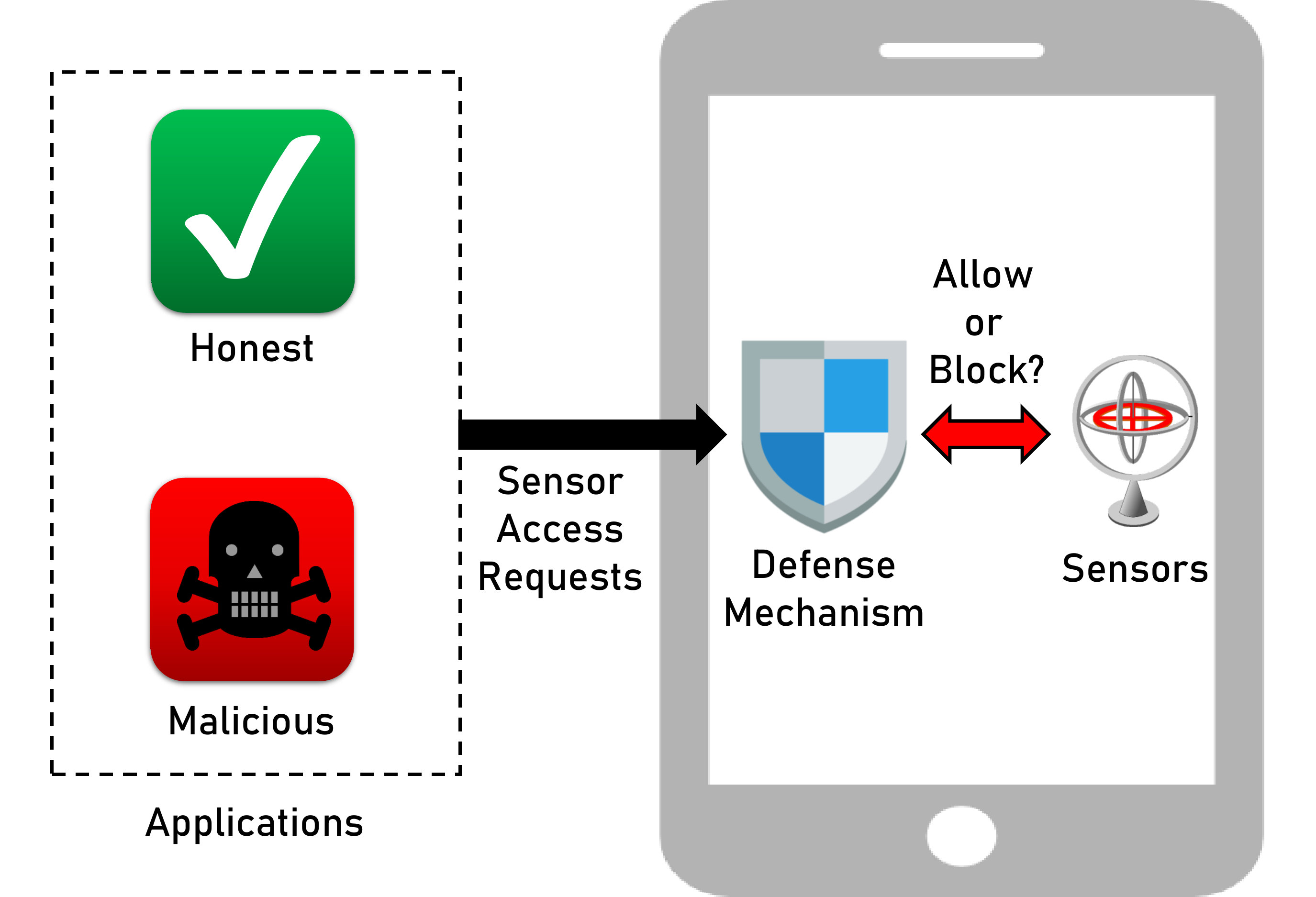}
}
%

\centering
\subfloat[Extensive form of the Sensor Access Signaling Game $\mathbb{G}_D=<\mathbb{P}, \mathbb{T}, \mathbb{S}, \mathbb{A}, \mathbb{U}, \theta, (p,q)>$.\label{fig:gamemodel}]{
\includegraphics[width=.80\textwidth]{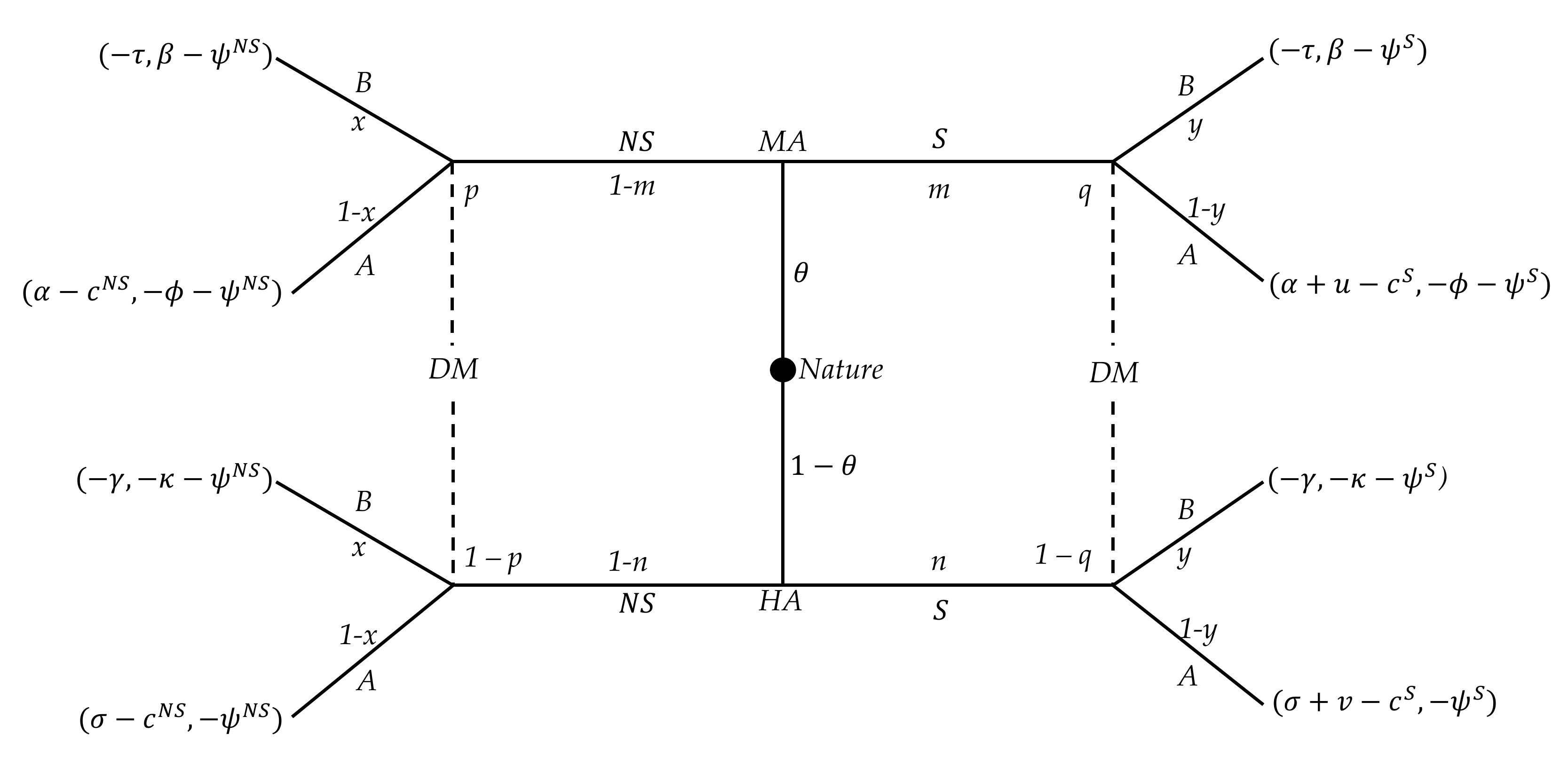}
}
\caption{Overview of the system and game models.}
\vspace{-0.2in}
\end{figure}

\noindent
\textbf{System Model.}
Our system (Figure \ref{fig:sysmodel}) comprises of two key entities residing on a user's (mobile) device. The first is \emph{applications} ($APP$) that utilize, and thus, need access to, data from zero-permission sensors. We consider two \emph{types} of applications: \textit{Honest ($HA$)} and \textit{Malicious ($MA$)}. Honest applications provide some useful service to the end-user with the help of zero-permission sensor data, while malicious applications would like to infer personal/private information about the user in the guise of offering some useful service. Both honest and malicious applications can request sensor data in a manner which may look normal/non-suspicious or suspicious (details next), regardless of their intentions or use-cases. 
The second entity is a sensor access regulator, which we refer to as the \textit{Defense Mechanism ($DM$)}. All sensor access requests (by all applications) must pass through and processed by the $DM$. The \emph{ideal} functionality that the $DM$ would like to achieve is to block sensor requests coming from $MA$s, while allowing requests from $HA$s. As noted earlier, the $DM$ itself does not know the type (i.e., honest or malicious) of application requesting sensor access - otherwise the job of the $DM$ is trivial. This is also a practical assumption as currently all applications can access these sensors without violating any system/user-defined policy (to clarify, there is currently no way to set access control policies for zero-permission sensors on most mobile platforms). As the $DM$ has no way of certainly knowing an application's true intentions (and thus, its type), it must rely on the received request (suspicious or non-suspicious, as described next) and its belief about the requesting application's type to determine whether it poses a threat to user privacy or not. 

\noindent
\textbf{Suspicious and Non-Suspicious Requests.}
Zero-permission sensor access requests by the applications (to the $DM$) can be classified as either \emph{suspicious} ($\mathcal{S}$) or \emph{non-suspicious} ($\mathcal{NS}$). Such a classification (generally, system-defined) can be accomplished using contextual information available to both the applications and the defense mechanism, such as, frequency, time, sampling rate, and relevance (according to the advertised type of service offered by the application) of these requests. Although there are several efforts in the literature in the direction of determining sensor over-privileges in mobile platforms \cite{felt2011android,hammad2017determination}, we abstract away this detail to keep our model general. 
We, however, assume that malicious applications are able to masquerade themselves perfectly as honest applications (in terms of the issued sensor requests), which is easy to accomplish when the target of these applications is zero-permission sensors. 

\noindent
\textbf{Other System Parameters.}
The strategic interactions between the (honest or malicious) $APP$ and $DM$ can be characterized using several system parameters which we summarize in Table \ref{tab:Symbols}. In addition to identifying these parameters, we also establish the relationship between these parameters by considering realistic network and system constraints as discussed next.
For example, if the cost of an application processing a successful $\mathcal{S}$ request (i.e., $c^{\mathcal{S}}$) or $\mathcal{NS}$ request (i.e., $c^{\mathcal{NS}}$) is expressed in terms of the CPU utilization (of the application), then it is clear that $c^{\mathcal{S}}\geq c^{\mathcal{NS}} $ because suspicious requests would usually solicit fine-grained (high sampling rate) sensor data compared to non-suspicious requests, thus requiring more processing time.
By a similar rationale, $\psi^{\mathcal{S}} \geq \psi^{\mathcal{NS}}$, where $\psi^{\mathcal{S}}$ and $\psi^{\mathcal{NS}}$ are the costs to a $DM$ (or the system) for processing a $\mathcal{S}$ or $\mathcal{NS}$ request, respectively. 
Now, the cost to the $HA$ in terms of loss in usability when its request is blocked by $DM$ (i.e., $\gamma$) and benefit for the $HA$ in terms of gain in usability when its request is allowed by the $DM$ (i.e., $\sigma$) are inversely proportional ($\gamma \propto 1/\sigma$).
Similarly, benefit to the $MA$ when it's request is allowed by $DM$ ($\alpha$) can be expressed in terms of monetary gains. An acute example would be if $MA$ is able to successfully infer user's banking credentials using sensor data \cite{Liu:2015:GBE:2810103.2813668,Wang:2016:FFY:2897845.2897847,maiti:2016,SabraMJ:2018}, and uses it for theft. A more clement example of monetary gain could be through selling contextual data (inferred from sensor data) to advertising companies, without user's consent. Accordingly, $MA$ is set back with a proportional cost ($\tau$) if its request is rejected by $DM$, i.e., $\alpha \propto \tau$. On the other hand, $DM$'s cost of allowing a $MA$'s request ($\phi$) versus benefit to the $DM$ for blocking $MA$'s request ($\beta$) are also inversely proportional ($\phi \propto 1/\beta$). $DM$'s cost of allowing a $MA$'s request is essentially borne by the user, but since the $DM$ is working in the best interest of the user, we combine their costs and benefits. Consequently, in case $DM$ blocks an $HA$'s request, it incurs a cost ($\kappa$) representing loss of utility/usability for the user. Lastly, we also capture the \emph{difference in benefits} for $MA$ and $HA$, in case they send out a $\mathcal{S}$ versus $\mathcal{NS}$ request, as $u$ and $v$, respectively. In essence, $u$ denotes the gain in benefit due to $MA$'s better inference accuracy caused by sensor data obtained from $\mathcal{S}$, and $v$ denotes the improvement of $HA$'s utility/usability due to sensor data obtained from $\mathcal{S}$. We also assume that these different (discrete) costs and benefits are appropriately scaled and normalized such that their absolute values lie in the same range of real values. 
Next, we outline the signaling game formulation to capture the strategic interaction between the mobile applications (requesting zero-permission sensor access) and the defense mechanism (attempting to regulating these requests).

\begin{table}[]
\scriptsize
\def\arraystretch{1.15}
\caption{System entities and parameters.}
\begin{center}
\begin{tabular}{c|L{11cm}} \toprule
	\textbf{Symbol}  		& \textbf{Definition}   \\  \rowcolor[gray]{.9} \hline
	$DM$				& Defense Mechanism \\
	$HA$				& Honest Application \\ \rowcolor[gray]{.9}
	$MA$				& Malicious Application \\
	$\theta$			& Probability that Nature selects $MA$ \\ \rowcolor[gray]{.9}
	$\mathcal{S}$		& Suspicious sensor request  \\ 
	 $\mathcal{NS}$		& Non-suspicious sensor request \\ \rowcolor[gray]{.9}
	$q$				& Belief probability of the $DM$ that the requester is of type $MA$ on receiving a $\mathcal{S}$ request \\
	$p$				& Belief probability of the $DM$ that the requester is of type $MA$ on receiving a $\mathcal{NS}$ request \\ \rowcolor[gray]{.9}
	$B$				& $DM$ response to block a sender request \\ 
	$A$				& $DM$ response to allow a sender request \\\rowcolor[gray]{.9}
	$c^{\mathcal{S}}$		& Cost of an application processing a successful $\mathcal{S}$ request\\
	$c^{\mathcal{NS}}$	& Cost of an application processing a successful $\mathcal{NS}$ request\\\rowcolor[gray]{.9}
	$\gamma$			& Cost to the $HA$ when its request is blocked by $DM$\\
	${\psi}^{\mathcal{S}}$	& Cost of a $DM$ processing a $\mathcal{S}$ request\\\rowcolor[gray]{.9}
	${\psi}^{\mathcal{NS}}$	& Cost of a $DM$ processing a $\mathcal{NS}$ request\\
	$\phi$				& Cost to the $DM$ when $MA$'s request is allowed \\\rowcolor[gray]{.9}
	$\tau$			& Cost to the $MA$ when its request is blocked by the $DM$\\
	$\kappa$			& Cost to the $DM$ when $HA$'s request is blocked \\\rowcolor[gray]{.9}
	$\alpha$			& Benefit to the $MA$ when its request is allowed by the $DM$\\
	$\beta$			& Benefit to the $DM$ for blocking $MA$'s request \\\rowcolor[gray]{.9}
	$\sigma$			& Benefit to the $HA$ when its request is allowed by the $DM$\\
	$u$				& Benefit difference to $MA$ for sending $\mathcal{S}$ instead of  $\mathcal{NS}$ \\\rowcolor[gray]{.9}
	$v$				& Benefit difference to $HA$ for sending $\mathcal{S}$ instead of  $\mathcal{NS}$\\
	\bottomrule
\end{tabular}
\end{center}
\label{tab:Symbols}
\vspace{-0.3in}
\end{table}

\noindent
\textbf{Game Model.} A classical signaling game \cite{cho1987signaling} is a sequential two-player incomplete information game in which \textit{Nature} starts the game by choosing the \emph{type} of the first player or \textit{player 1}. Player 1 is the more informed out of the two players since it knows the choice of \textit{Nature} and can send \emph{signals} to the less informed player, i.e., \textit{player 2}. Player 2 is uncertain about the type of player 1, and must base its strategic response solely based on the signal received from player 1. In other words, player 2 must decide its best response to player 1's signal without any knowledge about the type of player 1. Both players receive some utility (payoff) depending on the signal and type of player 1 and the response by player 2 (to player 1's signal). 
Both the players are assumed to be rational and are interested in solely maximizing their individual payoffs.  

Given the above generic description of the signaling game, let us briefly describe how our zero-permission sensor access scenario naturally lends itself as a single-stage signaling game. We refer to this game as the \emph{Sensor Access Signaling Game} and is formally represented as $\mathbb{G}_D=<\mathbb{P}, \mathbb{T}, \mathbb{S}, \mathbb{A}, \mathbb{U}, \theta, (p,q)>$, where $\mathbb{P}$ is the set of players, $\mathbb{T}$ is the set of player 1 types, $\mathbb{S}$ is the set of player 1 signals, $\mathbb{A}$ is the set of player 2 actions, $\mathbb{U}$ is the \emph{payoff/utility function}, $\theta$ is the \emph{Nature's probability distribution function}, and $(p,q)$ are player 2's \emph{belief functions} about player 1's type. Each sensor access request by an application can be modeled as a single stage of the above signaling game. In each such stage, $\mathbb{P}$ contains two players, i.e., $APP$ which is player 1 and the $DM$ which is player 2. As there are two types of applications (or player 1), i.e., honest ($HA$) and malicious ($MA$), $\mathbb{T} \equiv \{HA, MA\}$. As applications can send two types of signals (or requests), i.e., suspicious ($\mathcal{S}$) and non-suspicious ($\mathcal{NS}$),  $\mathbb{S}\equiv\{\mathcal{S},\mathcal{NS}\}$. As the $DM$ (or player 2) takes two types of actions depending on the received signal from player 1, i.e., Allow ($A$) or Block ($B$), $\mathbb{A} \equiv \{A,B\}$. The utility function $\mathbb{U}:\mathbb{T} \times \mathbb{S} \times \mathbb{A} \to (\mathbb{R},\mathbb{R})$ assigns a real-valued payoff to each player (at the end of the stage) based on the benefit received and the cost borne by each player, and is outlined in the extensive form of the game depicted in Figure \ref{fig:gamemodel}. The first utility in the pair is the $APP$'s utility denoted as $U_{APP}$, while the second utility in the pair is the $DM$'s utility denoted as $U_{DM}$.

Lastly, let $\Gamma_{APP} = \{\mu_{APP}|\forall t_i \in \mathbb{T}, \sum_{\lambda \in \mathbb{S}}  \mu_{APP}(\lambda|t_i); \forall t_i \in \mathbb{T}\}$ and $\Gamma_{DM} = \{\mu_{DM}|\forall \lambda \in \mathbb{S}, \sum_{a \in \mathbb{A}}  \mu_{DM}(a|\lambda); \forall \lambda \in \mathbb{S}\}$ be the strategy spaces for $APP$ and $DM$, respectively. A strategy $\mu_{APP}$ for the $APP$ and $\mu_{DM}$ for the $DM$ can be either \emph{pure} or \emph{mixed}, as identified by parameters $m$, $n$, $y$ and $x$ in Figure \ref{fig:gamemodel}. For pure strategies $m, n, y, x \in \{0,1\}$, while for mixed strategies $0<m, n, y, x<1$. Moreover, let us represent each of the $DM$'s belief functions by conditional (posterior) probability distributions as $q=Pr(MA|\mathcal{S})$ and $p=Pr(MA|\mathcal{NS})$, which also imply that $1-q = Pr(HA|\mathcal{S})$ and $1-p=Pr(HA|\mathcal{NS})$. 

Now, let's characterize the set of equilibrium strategies in $\mathbb{G}_D$, i.e., a set of strategy pairs that are mutual best responses to each other and no player has any incentive to move away from their strategy in that pair.
In order to determine mutual best responses, we need to evaluate the actions (or strategies) of each player at each \emph{information set} of the game. $APP$'s information set comprises of a single decision point (i.e., to select a signal $\lambda \in \{ \mathcal{S}, \mathcal{NS} \}$) after Nature makes its selection of the type ($HA$ or $MA$) and reveals it to $APP$. 
$DM$'s information set, on the other hand, comprises of two decision points because of its incomplete information about the type of $APP$ chosen by Nature. Thus, $DM$'s strategy is to select an action $a \in \{A, B\}$ depending on its belief $Pr(t_{i}|\lambda)$  about the type $t_{i} \in \mathbb{T}$ of $APP$ in that information set. 
Moreover, for each $\lambda \in \{\mathcal{S},\mathcal{NS}\}$, $\sum_{t_{i}} Pr(t_{i}|\lambda) = 1$.

Our goal is to determine the existence of \textit{Perfect Bayesian Nash Equilibria (or PBNE)} in $\mathbb{G}_D$, where strategies are combined with beliefs to determine the mutual best responses of each player at the end of each stage. A PBNE of the Sensor Access Signaling Game $\mathbb{G}_D$ is a strategy profile $\mu^{*} = (\mu^{*}_{APP},\mu^{*}_{DM})$  and posterior probabilities (or beliefs of the $DM$) $Pr(t_{i}|\lambda)$ such that:
{\scriptsize
$$\mu^{*}_{APP} \in argmax_{\mu_{APP} \in \Gamma_{APP}} U_{APP}(\mu_{APP}, \mu^{*}_{DM}, t_i); \forall t_i \in \mathbb{T}$$
}
where, $U_{APP}(.)$ is the utility or payoff of $APP$ for a particular pure or mixed strategy $\mu_{APP}$ against $DM$'s best response to it, when the type $t_i$ selected by Nature, and, $\forall \lambda \in \mathbb{S}=\{\mathcal{S},\mathcal{NS}\}$ such that:
{\scriptsize
$$\mu^{*}_{DM} \in argmax_{\mu_{DM} \in \Gamma_{DM}} \sum_{t_i \in \mathbb{T}} Pr(t_i|\lambda)\ U_{DM}(\lambda, \mu_{DM}, t_i)$$
}
where, $U_{DM}(.)$ is the payoff of $DM$ for a particular pure or mixed strategy $\mu_{DM}$ against the signal ($\lambda$) received from the $APP$, when the type $t_i$ selected by Nature. Moreover, the $DM$'s belief $Pr(t_{i}|\lambda)$ about the $APP$'s type given a received signal $\lambda$ should satisfy Bayes' theorem, i.e., 
{\scriptsize
$$Pr(t_{i}|\lambda) = \frac{Pr(\lambda|t_{i})Pr(t_{i})}{Pr(\lambda)} = \frac{\mu_{APP}(\lambda|t_{i})Pr(t_{i})}{Pr(\lambda)}$$
}
%
%
%
  
Four categories of PBNE can exist for a signaling game such as $\mathbb{G}_D$:
\begin{itemize}[leftmargin=*]
\item \textbf{Separating PBNE:} This category comprises of strategy profiles where player 1 or $APP$ of different types dominantly send different or contrasting types of signals $\lambda \in \{\mathcal{S},\mathcal{NS}\}$. This allows $DM$ to infer $APP$'s type with certainty. For instance, in a separating strategy profile $\{(\mathcal{S},\mathcal{NS}),\mu^{*}_{DM}\}$, $APP$ of $MA$ type always selects $\mathcal{S}$ (i.e., $m=1$) while $HA$ always selects the $\mathcal{NS}$ (i.e., $n=0$).
\item \textbf{Pooling PBNE:} This category comprises of strategy profiles where player 1 or $APP$ of different types dominantly send the same type of signal $\lambda$. Here $DM$ cannot infer $APP$'s type with certainty, but needs to update its belief (about $APP$'s type) based on the observed $\lambda$. For instance, in a pooling strategy profile $\{(\mathcal{S},\mathcal{S}),\mu^{*}_{DM}\}$, both $MA$ and $HA$ types always select $\mathcal{S}$ (i.e., $m,n=1$).
\item \textbf{Hybrid PBNE:} This category comprises of strategy profiles where one player 1 or $APP$ type dominantly sends one type of signal, but the other type randomizes its sent signal. For instance, in a hybrid strategy profile $\{(\mathcal{S},(\mathcal{S},\mathcal{NS})),\linebreak\mu^{*}_{DM}\}$, $MA$ always selects $\mathcal{S}$  (i.e., $m=1$), whereas $HA$ randomizes between $\mathcal{S}$ and $\mathcal{NS}$ (i.e., $0<n<1$).
\item \textbf{Mixed PBNE:} Finally, this equilibrium comprises of strategy profiles where all player 1 or $APP$ types send signals $\lambda$ only in a probabilistic fashion (i.e., $0<m,n<1$).
\end{itemize}


\section{Game Analysis}
\label{sec:gameanalysis}

In this section, we find the PBNE for the sensor access signaling game $\mathbb{G}_D$. We begin by evaluating the existence of pure strategy equilibria (i.e., separating, pooling and hybrid), including conditions and regimes for achieving these equilibria. Following that we determine the mixed strategy equilibria for $\mathbb{G}_D$.

\begin{theorem}\label{thm:separating}
There does not exist a separating equilibrium in the game $\mathbb{G}_D$. 
\end{theorem}

\begin{proof}
There can be two possible separating strategy profiles for $APP$: $(\mathcal{S},NS)$ and $(\mathcal{NS}, S)$. First, let us analyze the existence of an equilibrium on $(\mathcal{S},\mathcal{NS})$, which means $MA$ (malicious type) always selects $\mathcal{S}$ (i.e., $m=1$) while $HA$ (honest type) always selects $\mathcal{NS}$ (i.e., $n=0$). $DM$'s beliefs for the  can be calculated using Bayes' theorem as follows:
{\scriptsize
\begin{eqnarray}
Pr(MA|\mathcal{S}) = q & = & \frac{Pr(\mathcal{S}|MA) \times Pr(MA)}{Pr(\mathcal{S})} = \frac{Pr(\mathcal{S}|MA) \times Pr(MA)}{Pr(\mathcal{S}|MA) \times Pr(MA) + Pr(\mathcal{S}|HA) \times Pr(HA)}\nonumber\\
& = & \frac{m \times \theta}{m \times \theta + n \times (1 - \theta)} = \frac{1 \times \theta}{1 \times \theta + 0 \times (1 - \theta)} = 1\nonumber
\end{eqnarray}
}
Therefore, $Pr(HA|\mathcal{S})=1-q=0$. Similarly, we can show that $p=0$, and $1-p=1$. With these beliefs, the best response of $DM$ can be calculated as follows. The $DM$'s expected utility/payoff ($EU_{DM}$) from playing $B$ or $A$ if $MA$ or $HA$ selects $\mathcal{S}$ are:
{\scriptsize
$$EU_{DM}(B,\mathcal{S}) = 1 \times (\beta - \psi^\mathcal{S}) + 0 \times (-\kappa - \psi^\mathcal{S}) = \beta - \psi^\mathcal{S}$$
$$EU_{DM}(A,\mathcal{S}) = 1 \times (-\phi - \psi^\mathcal{S}) +  0 \times (- \psi^\mathcal{S}) = -\phi - \psi^\mathcal{S}$$
}
As $EU_{DM}(B,\mathcal{S}) > EU_{DM}(A,\mathcal{S})$, the $DM$'s best response in this case is to play Block, i.e., $BR_{DM}(\mathcal{S}) = B$. Similarly, the $DM$'s expected utility/payoff from playing $B$ or $A$ if $MA$ or $HA$ selects $\mathcal{NS}$ are:
{\scriptsize
$$EU_{DM}(B,\mathcal{NS}) = 0 \times (\beta - \psi^\mathcal{NS}) + 1 \times (-\kappa - \psi^\mathcal{NS}) = -\kappa - \psi^\mathcal{NS}$$
$$EU_{DM}(A,\mathcal{NS}) = 0 \times (-\phi - \psi^\mathcal{NS}) +  1 \times (-\psi^\mathcal{NS}) = -\psi^\mathcal{S}$$
}
In this case, as $EU_{DM}(B,\mathcal{NS}) < EU_{DM}(A,\mathcal{NS})$, the $DM$'s best response is to play Allow, i.e., $BR_{DM}(\mathcal{NS}) = A$. In summary, if $MA$ or $HA$ plays $\mathcal{S}$ then $DM$'s best response is $B$, and if $MA$ or $HA$ plays $\mathcal{NS}$ then $DM$'s best response is $A$. 

\textit{Check for Equilibrium:} $HA$ and $MA$ will follow the strategy along the equilibrium path as long as the payoff along that path is higher than the payoff it will get if it deviates. There can be two scenarios: first if the $MA$ deviates and plays $\mathcal{NS}$ and second if the $HA$ deviates and plays $\mathcal{S}$. Let us first analyze the case where $MA$ deviates and plays $\mathcal{NS}$. The $DM$'s beliefs do not change, and so, if it sees $MA$ or $HA$ playing $\mathcal{NS}$, it will still always respond with it's best response, i.e., $A$. $MA$ will receive a payoff of $-\tau$ if it plays $\mathcal{S}$ and will receive a payoff of $\alpha-c^\mathcal{NS}$ if it plays $\mathcal{NS}$. Thus, $MA$ has an incentive to deviate from the equilibrium path. Although it can be shown that $HA$ does not have an incentive to deviate, equilibrium does not exist in this case because at least one $APP$ (player 1) type has an incentive to deviate.

Next, let us analyze the existence of a separating equilibrium on $(\mathcal{NS},\mathcal{S})$, which means $MA$ always selects $\mathcal{NS}$ (i.e., $m=0$) and $HA$ always selects $\mathcal{S}$ (i.e., $n=1$). As before, the belief functions for the $DM$ can be calculated as:
{\scriptsize
$$Pr(MA|\mathcal{NS}) = p = \frac{Pr(\mathcal{NS}|MA) \times Pr(MA)}{Pr(\mathcal{NS})} = \frac{1 \times \theta}{1 \times \theta + 0 \times (1 - \theta)} = 1$$
}
Therefore, $Pr(HA|\mathcal{NS})=1-p=0$. Similarly, we can also show that $q=0$ and $1-q=1$. Thus, the $DM$'s expected utility/payoff from playing $B$ or $A$ if $MA$ or $HA$ selects $\mathcal{S}$ are:
{\scriptsize
$$EU_{DM}(B,\mathcal{S}) = 0 \times (\beta - \psi^\mathcal{S}) + 1 \times (-\kappa - \psi^\mathcal{S}) = -\kappa - \psi^\mathcal{S}$$
$$EU_{DM}(A,\mathcal{S}) = 0 \times (-\phi - \psi^\mathcal{S}) +  1 \times (-\psi^\mathcal{S}) = - \psi^\mathcal{S}$$
}
In this case, as $EU_{DM}(B,\mathcal{S}) < EU_{DM}(A,\mathcal{S})$, the $DM$'s best response is to play Allow, i.e., $BR_{DM}(\mathcal{S}) = A$. And, $DM$'s expected utility from playing $B$ or $A$ if $MA$ or $HA$ selects $\mathcal{NS}$ are:
{\scriptsize
$$EU_{DM}(B,\mathcal{NS}) = 1 \times (\beta - \psi^\mathcal{NS}) + 0 \times (-\kappa - \psi^\mathcal{NS}) = \beta - \psi^\mathcal{NS}$$
$$EU_{DM}(A,\mathcal{NS}) = 1 \times (-\phi - \psi^\mathcal{NS}) +  0 \times (-\psi^\mathcal{NS}) = -\phi - \psi^\mathcal{NS}$$
}
As $EU_{DM}(B,\mathcal{NS})>EU_{DM}(A,\mathcal{NS})$, in this case the $DM$'s best response is to Block, i.e., $BR_{DM}(\mathcal{NS}) = B$. In summary, if $MA$ or $HA$ plays $\mathcal{S}$, then $DM$'s best response is $A$ and if $MA$ or $HA$ plays $\mathcal{NS}$, then $DM$'s best response is $B$. 

\textit{Check for Equilibrium:} If $MA$ deviates and plays $\mathcal{S}$, $DM$ will respond with it's best response $A$. As a result, $MA$ will receive a payoff of $-\tau$ if it plays $\mathcal{NS}$ and will receive a payoff of $\alpha+u-c^\mathcal{S}$ if it plays $\mathcal{S}$. Thus, $MA$ has an incentive to deviate from the equilibrium path. Again, although it can be shown that $HA$ does not have an incentive to deviate, equilibrium does not exist in this case either because at least one $APP$ (player 1) type has incentive to deviate.

\textit{Thus, neither of the separating strategy profiles $\{(\mathcal{S},\mathcal{NS}), (B, A), \linebreak p, q\}$ and $\{(\mathcal{NS},\mathcal{S}), (A, B), p, q\}$ is a PBNE.}
\end{proof}

\begin{theorem}\label{thm:pooling-s-s}
There exists a pooling equilibrium on $APP$ strategy of $(\mathcal{S},\mathcal{S})$ in the game $\mathbb{G}_D$.
\end{theorem}

\begin{proof}
An $APP$ strategy profile ($\mathcal{S},\mathcal{S}$) means both $MA$ and $HA$ types always select $\mathcal{S}$ (i.e., $m,n=1$). $DM$'s beliefs in this strategy profile can be calculated as:
{\scriptsize
$$Pr(MA|\mathcal{S}) = q = \frac{Pr(\mathcal{S}|MA) \times Pr(MA)}{Pr(\mathcal{S})} = \frac{1 \times \theta}{1 \times \theta + 1 \times (1 - \theta)} = \theta$$
}
Therefore, $Pr(HA|\mathcal{S})=1-q=1-\theta$. Accordingly, expected payoff for $DM$ from playing $B$ or $A$ if either $MA$ or $HA$ selects $\mathcal{S}$ are:
{\scriptsize
\begin{eqnarray}
EU_{DM}(B,\mathcal{S}) & = & \theta \times (\beta - \psi^\mathcal{S}) + (1 - \theta)  \times (-\kappa - \psi^\mathcal{S})\nonumber\\
& = & \theta(\beta + \kappa) - \kappa - \psi^\mathcal{S}\nonumber\\
EU_{DM}(A,\mathcal{S}) & = & \theta \times (-\phi - \psi^\mathcal{S}) + (1 - \theta) \times (- \psi^\mathcal{S})\nonumber\\
& = & -\phi \times \theta - \psi^\mathcal{S}\nonumber
\end{eqnarray}
}
Now, $DM$'s best response to the $APP$'s pooling strategy of ($\mathcal{S},\mathcal{S}$) would be to select $B$ (over $A$) if and only if the following condition holds:
{\scriptsize
$$\theta(\beta + \kappa) - \kappa - \psi^\mathcal{S} \geq - \phi \times \theta - \psi^\mathcal{S} \equiv \theta \geq \frac{\kappa}{\beta + \kappa + \phi}$$ 
}
To analyze the existence of an equilibrium at the $APP$'s strategy of ($\mathcal{S},\mathcal{S}$), given the $DM$'s best response, we must check if $APP$ of either type ($MA$ or $HA$) has an incentive to deviate and play $\mathcal{NS}$. Here, if $HA$ or $MA$ deviate and play $\mathcal{NS}$ and $DM$ chooses $A$, $HA$ gains a payoff of $\sigma - c^\mathcal{NS}$ compared to $- \gamma$ if it plays $\mathcal{S}$, while $MA$ gains a payoff of $\alpha - c^\mathcal{NS}$ compared to $- \tau$ if it plays $\mathcal{S}$. Thus, in this case both $HA$ and $MA$ have an incentive to deviate and play $\mathcal{NS}$ and there is no equilibrium. Here, if $HA$ or $MA$ deviate and play $\mathcal{NS}$ and $DM$ chooses $B$, $HA$ will receive a payoff of $-\gamma$, same as if it plays $\mathcal{S}$, while $MA$ will receive a payoff of $-\tau$, same as if it plays $\mathcal{S}$. Thus, in this case, both $HA$ and $MA$ do not have any incentive to switch to $\mathcal{NS}$ and an equilibrium exists. In summary, an equilibrium on the $APP$'s pooling strategy of ($\mathcal{S},\mathcal{S}$) exists when $\theta \geq \frac{\kappa}{\beta + \kappa + \phi}$. 

Inversely, the $DM$'s best response to $APP$'s pooling strategy of ($\mathcal{S},\mathcal{S}$) would be to select $A$ (over $B$) if and only if the following holds:
{\scriptsize
$$\theta(\beta + \kappa) - \kappa - \psi^\mathcal{S}  \leq - \phi \times \theta - \psi^\mathcal{S} \equiv \theta \leq \frac{\kappa}{\beta + \kappa + \phi}$$
}
Here, if $HA$ or $MA$ deviate and play $\mathcal{NS}$ and $DM$ chooses $A$, $HA$ will receive a payoff of $\sigma - c^\mathcal{NS}$ if it plays $\mathcal{NS}$ and will receive a payoff of $\sigma + v - c^\mathcal{S}$ if it plays $\mathcal{S}$. On the other hand, $MA$ will receive a payoff of $\alpha - c^\mathcal{NS}$ if it plays $\mathcal{NS}$ and will receive a payoff of $\alpha + u - c^\mathcal{S}$ if it plays $\mathcal{S}$. Thus, in this case, there will be a pooling equilibrium if and only if:            
{\scriptsize           
$$\sigma + v - c^\mathcal{S} \geq \sigma - c^\mathcal{NS} \equiv v \geq c^\mathcal{S} -  c^\mathcal{NS},\ and$$                
$$\alpha + u - c^\mathcal{S} \geq \alpha - c^\mathcal{NS} \equiv u \geq c^\mathcal{S} - c^\mathcal{NS}$$
}
Here, if $HA$ or $MA$ deviate and play $\mathcal{NS}$ and $DM$ chooses $B$, $HA$ will receive a payoff $- \gamma$ compared to $\sigma + v - c^\mathcal{S}$ if it plays $\mathcal{S}$, while $MA$ will receive a payoff of $-\tau$ compared to $\alpha + u - c^\mathcal{S}$ if it plays $\mathcal{S}$. Thus, in this particular case, $HA$ and $MA$ do not have any incentive to deviate as well. In summary, an equilibrium on $APP$'s pooling strategy of ($\mathcal{S},\mathcal{S}$) also exists when $\theta \leq \frac{\kappa}{\beta + \kappa + \phi}$.
\end{proof}

\begin{theorem}\label{thm:pooling-ns-ns}
There exists a pooling equilibrium on $APP$ strategy of $(\mathcal{NS},\mathcal{NS})$ in the game $\mathbb{G}_D$.
\end{theorem}

\begin{proof}
An $APP$ strategy profile ($\mathcal{NS},\mathcal{NS}$) means that both $MA$ and $HA$ types always select $\mathcal{NS}$ (i.e., $m,n=0$). $DM$'s beliefs in this strategy profile can thus be calculated as:
{\scriptsize
$$Pr(MA|\mathcal{NS}) = p = \frac{Pr(\mathcal{NS}|MA) \times Pr(MA)}{Pr(\mathcal{NS})} = \frac{1 \times \theta}{1 \times \theta + 1 \times (1 - \theta)} =  \theta$$
}
Therefore, $Pr(HA|\mathcal{NS})=1-p=1-\theta$. Accordingly, expected payoff for $DM$ from playing $B$ or $A$ if either $MA$ or $HA$ selects $\mathcal{S}$ are:
{\scriptsize
\begin{eqnarray}
EU_{DM}(B,\mathcal{NS}) & = & \theta \times (\beta - \psi^\mathcal{NS}) + (1 - \theta) \times (-\kappa - \psi^\mathcal{NS})\nonumber\\
& = & \theta( \beta + \kappa) -\kappa -\psi^\mathcal{S}\nonumber\\
EU_{DM}(A,\mathcal{NS}) & = & \theta \times (-\phi - \psi^\mathcal{NS}) +  (1 - \theta) \times (-\psi^\mathcal{NS})\nonumber\\
& = & -\phi \times \theta -\psi^\mathcal{NS} \nonumber
\end{eqnarray}
}
Now, $DM$'s best response to $APP$'s pooling strategy of ($\mathcal{NS},\mathcal{NS}$) would be to select  $B$ (over $A$) if and only if the following holds:
{\scriptsize
$$\theta( \beta + \kappa) - \kappa - \psi^\mathcal{NS} \geq -\phi \times \theta - \psi^\mathcal{NS} \equiv \theta \geq \frac{\kappa}{\beta + \kappa + \phi}$$ 
}
To analyze the existence of an equilibrium at the $APP$'s strategy of ($\mathcal{NS},\mathcal{NS}$), given the $DM$'s best response, we must check if $APP$ of either type ($MA$ or $HA$) has an incentive to deviate and play $\mathcal{S}$. Here, if $HA$ or $MA$ deviate and play $\mathcal{S}$ and $DM$ play $A$, $HA$ will gain a payoff of $\sigma + v - c^\mathcal{S}$ compared to $- \gamma$ if it plays $\mathcal{NS}$, while $MA$ will gain a payoff of $\alpha + u - c^\mathcal{S}$ compared to $- \tau$ if it plays $\mathcal{NS}$. Thus, in this case, both $HA$ and $MA$ have an incentive to deviate and play $\mathcal{S}$ and there is no equilibrium. Here, if $HA$ or $MA$ deviate and play $\mathcal{S}$ and $DM$ chooses $B$, $HA$ will receive a payoff of $-\gamma$, same as if it plays $\mathcal{NS}$ and $MA$ will receive a payoff of $-\tau$, same as if it plays $\mathcal{NS}$.
Thus, in this case, both $HA$ and $MA$ do not have any incentive to switch to $\mathcal{S}$ and an equilibrium exists. In summary, an equilibrium on the $APP$'s pooling strategy of ($\mathcal{NS},\mathcal{NS}$) exists when $\theta \geq \frac{\kappa}{\beta + \kappa + \phi}$. 

Inversely, the $DM$'s best response to the $APP$'s pooling strategy of ($\mathcal{NS},\mathcal{NS}$) would be to select $A$ (over $B$) if and only if the following condition holds:
{\scriptsize
$$\theta( \beta + \kappa) - \kappa - \psi^\mathcal{NS}  \leq -\phi \times \theta - \psi^\mathcal{NS} \equiv \theta \leq \frac{\kappa}{\beta + \kappa + \phi}$$
}

Here, if $HA$ or $MA$ deviate and play $\mathcal{S}$ and $DM$ chooses $A$, $HA$ will receive a payoff of $\sigma - c^\mathcal{NS}$ if it plays $\mathcal{NS}$ and will receive a payoff of $\sigma + v - c^\mathcal{S}$ if it plays $\mathcal{S}$. On the other hand, $MA$ will receive a payoff of $\alpha - c^\mathcal{NS}$ if it plays $\mathcal{NS}$ and will receive a payoff of $\alpha + u - c^\mathcal{S}$ if it plays $\mathcal{S}$. Thus, in this case, there will be a pooling equilibrium, if and only if:
{\scriptsize
$$\sigma + v - c^\mathcal{S} \leq \sigma - c^\mathcal{NS} \equiv v \leq c^\mathcal{S} -  c^\mathcal{NS},\ and$$
$$\alpha + u - c^\mathcal{S} \leq \alpha - c^\mathcal{NS} \equiv u \leq c^\mathcal{S} -  c^\mathcal{NS}$$
}     
Here, if $MA$ or $HA$ deviate and play $\mathcal{S}$ and $DM$ chooses $B$, $HA$ will receive a payoff of $-\gamma$, compared to $\sigma - c^\mathcal{NS}$ if it plays $\mathcal{NS}$, while $MA$ will receive a payoff of $-\tau$ compared to $\alpha - c^\mathcal{NS}$ if it plays $\mathcal{NS}$. Thus, in this particular case, $HA$ and $MA$ do not have any incentive to deviate as well. In summary, an equilibrium on the $APP$'s pooling strategy of ($\mathcal{NS},\mathcal{NS}$) also exists when $\theta \leq \frac{\kappa}{\beta + \kappa + \phi}$.   
\end{proof}

\begin{theorem}\label{thm:hybrid-s-s-ns}
There exists a hybrid equilibrium on the $APP$ strategy profile $(\mathcal{S},\linebreak(\mathcal{S},\mathcal{NS}))$ in game $\mathbb{G}_D$.
\end{theorem}

\begin{proof}
An $APP$ strategy profile $(\mathcal{S},(\mathcal{S},\mathcal{NS}))$ means that $MA$ always selects $\mathcal{S}$ (i.e., $m=1$), whereas $HA$ selects $\mathcal{S}$ with some probability $n$ and $\mathcal{NS}$ with probability $1-n$ where ($0<n<1$). $DM$'s beliefs in this strategy profile can thus be calculated as:
{\scriptsize
$$Pr(MA|\mathcal{S}) = q = \frac{Pr(\mathcal{S}|MA) \times Pr(MA)}{Pr(\mathcal{S})} = \frac{1 \times \theta}{1 \times \theta + n \times (1 - \theta)} = \frac{\theta}{\theta(1-n) + n}$$
$$Pr(MA|\mathcal{NS}) = p = \frac{Pr(\mathcal{NS}|MA) \times Pr(MA)}{Pr(\mathcal{NS})} = \frac{0 \times \theta}{0 \times \theta + (1-n) \times (1 - \theta)} = 0$$
}
Now, let's compute the $DM$'s best response for each of the strategies $\mathcal{S}$ and $\mathcal{NS}$ of $APP$. In order to determine that, we need to first compute the expected utilities/payoffs obtained by $DM$ for playing $B$ or $A$ if $APP$ ($MA$ or $HA$) selects $\mathcal{NS}$ or $\mathcal{S}$, which is given by:
{\scriptsize
$$EU_{DM}(B,\mathcal{NS}) = p \times (\beta -\psi^\mathcal{NS}) + (1-p) \times (-\kappa -\psi^\mathcal{NS}) = -\kappa -\psi^\mathcal{NS}$$
$$EU_{DM}(A,\mathcal{NS}) = p \times (-\phi -\psi^\mathcal{NS}) + (1-p) \times (-\psi^\mathcal{NS}) = -\psi^\mathcal{NS}$$
$$EU_{DM}(B,\mathcal{S}) = q \times (\beta -\psi^\mathcal{S}) + (1-q) \times (-\kappa -\psi^\mathcal{S})$$
$$EU_{DM}(A,\mathcal{S}) = q \times (-\phi -\psi^\mathcal{S}) + (1-q) \times (-\psi^\mathcal{S})$$
}
It is clear from these expected utilities obtained by the $DM$ in this strategy profile that it will always plays $A$ (i.e., $A$ always dominates $B$) when the $APP$ plays $\mathcal{NS}$. On the contrary, there are two possibilities in terms of the $DM$'s best response to an application's strategy of $\mathcal{S}$. The first possibility is for the $DM$ to always Block or $B$, i.e., $B$ would dominate $A$. This, however, holds only if the following is true: 
{\scriptsize
$$q(\beta - \psi^\mathcal{S}) + (1-q)(-\kappa - \psi^\mathcal{S}) \geq  q(-\phi - \psi^\mathcal{S}) +  (1-q)(- \psi^\mathcal{S}) \equiv q \geq \frac{(1-q)\kappa}{\beta + \phi}$$
}
Now, as $DM$ always plays $A$ for $\mathcal{NS}$, $HA$ has more incentive to play $\mathcal{NS}$ because it will gain $\sigma - c^\mathcal{NS}$ compared to $-\gamma$ if it plays $\mathcal{S}$. Also, $MA$ has more incentive to play $\mathcal{NS}$ since it will gain $\alpha - c^\mathcal{NS}$ compared to $-\tau$ if it plays $\mathcal{S}$. In other words, $APP$ is not indifferent between playing $\mathcal{S}$ and $\mathcal{NS}$ when $q \geq \frac{(1-q)\kappa}{\beta + \phi}$, and strongly prefers playing $\mathcal{NS}$. Thus, there is no hybrid equilibria at $(\mathcal{S},(\mathcal{S},\mathcal{NS}))$ when $q \geq \frac{(1-q)\kappa}{\beta + \phi}$.

The second possibility, in terms of the $DM$'s best response to an $APP$'s strategy of $\mathcal{S}$, is for the $DM$ to Accept or $A$ (i.e., $A$ dominates $B$) which is true if $q \leq \frac{(1-q)\kappa}{\beta + \phi}$. This combined with the fact that the $DM$ always plays $A$ for $\mathcal{NS}$, it is clear that when $q \leq \frac{(1-q)\kappa}{\beta + \phi}$, $DM$ invariantly plays $A$ for both the $\mathcal{S}$ and $\mathcal{NS}$ strategies of the $APP$. In this case, if $MA$ deviates and plays $\mathcal{NS}$ it will gain $\alpha - c^\mathcal{NS}$ compared to $\alpha + u - c^\mathcal{S}$ if it plays $\mathcal{S}$.
Similarly, $HA$ will gain $\sigma - c^\mathcal{NS}$ instead of $\sigma + v - c^\mathcal{S}$ if it plays $\mathcal{S}$. Therefore, in order to make $APP$ indifferent between playing $\mathcal{S}$ and $\mathcal{NS}$ so that a hybrid equilibrium can be achieved at $(\mathcal{S},(\mathcal{S},\mathcal{NS}))$, the following conditions must be satisfied:
{\scriptsize
$$\alpha - c^\mathcal{NS} \simeq \alpha + u - c^\mathcal{S} \equiv c^\mathcal{S} - c^\mathcal{NS} \simeq u$$
$$\sigma - c^\mathcal{NS} \simeq \sigma + v - c^\mathcal{S} \equiv c^\mathcal{S} - c^\mathcal{NS} \simeq v$$ 
}
In summary, a hybrid equilibrium is possible at $(\mathcal{S},(\mathcal{S},\mathcal{NS}))$ if and only if the above conditions hold.
\end{proof}

\begin{theorem}\label{thm:hybrid-ns-s-ns}
There exists a hybrid equilibrium on the $APP$ strategy profile $(\mathcal{NS},(\mathcal{S},\mathcal{NS}))$ in game $\mathbb{G}_D$.
\end{theorem}

\begin{proof}
An $APP$ strategy profile $(\mathcal{NS},(\mathcal{S},\mathcal{NS}))$ means that $MA$ always selects $\mathcal{NS}$ (i.e., $m=0$), whereas $HA$ selects $\mathcal{S}$ with some probability $n$ and $\mathcal{NS}$ with probability $1-n$ where ($0<n<1$). $DM$'s beliefs in this strategy profile can thus be calculated as:
{\scriptsize
$$Pr(MA|\mathcal{NS}) = p = \frac{Pr(\mathcal{NS}|MA) \times Pr(MA)}{Pr(\mathcal{NS})} = \frac{1 \times \theta}{1 \times \theta + (1 - n) \times (1 - \theta)} = \frac{\theta}{\theta + (1-n)(1-\theta)}$$
$$Pr(MA|\mathcal{S}) = q = \frac{Pr(\mathcal{S}|MA) \times Pr(MA)}{Pr(\mathcal{S})} = \frac{0 \times \theta}{0 \times \theta + n \times (1 - \theta)} = 0$$
}
Now, let's compute the $DM$'s best response for each of the strategies $\mathcal{S}$ and $\mathcal{NS}$ of $APP$. In order to determine that, we need to first compute the expected utilities/payoffs obtained by  $DM$ for playing $B$ or $A$ if $APP$ ($MA$ or $HA$) selects $\mathcal{S}$ or $\mathcal{NS}$, which is given by:
{\scriptsize
$$EU_{DM}(B,\mathcal{S}) = q \times (\beta -\psi^\mathcal{S}) + (1-q) \times (-\kappa -\psi^\mathcal{S}) = -\kappa - \psi^\mathcal{S}$$
$$EU_{DM}(A,\mathcal{S}) = q \times (- \phi -\psi^\mathcal{S}) +  (1-q) \times (-\psi^\mathcal{S}) = -\psi^\mathcal{S}$$
$$EU_{DM}(B,\mathcal{NS}) = p \times (\beta -\psi^\mathcal{NS}) + (1-p)  \times (-\kappa -\psi^\mathcal{NS})$$
$$EU_{DM}(A,\mathcal{NS}) = p \times (-\phi -\psi^\mathcal{NS}) +  (1-p) \times (-\psi^\mathcal{NS})$$
}
It is clear from these expected utilities obtained by the $DM$ in this strategy profile that it will always plays $A$ (i.e., $A$ always dominates $B$) when the $APP$ plays $\mathcal{S}$. On the contrary, there are two possibilities in terms of the $DM$'s best response to an $APP$'s strategy of $\mathcal{NS}$. The first possibility is for the $DM$ to always Block or $B$, i.e., $B$ would dominate $A$. This, however, holds only if the following is true: 
{\scriptsize
$$p(\beta -\psi^\mathcal{NS}) + (1-p)(-\kappa -\psi^\mathcal{NS}) \geq  p(-\phi -\psi^\mathcal{NS}) + (1-p)(-\psi^\mathcal{NS})$$ 
$$\equiv p \geq \frac{(1-p)\kappa}{\beta + \phi}$$
}

Now, as $DM$ always plays $A$ for $\mathcal{S}$, $HA$ has more incentive to play $\mathcal{S}$ because it will gain $\sigma + v - c^\mathcal{S}$ compared to $-\gamma$ if it plays $\mathcal{NS}$. Also, $MA$ has more incentive to play $\mathcal{S}$ since it will gain $\alpha + u- c^\mathcal{S}$ compared to $-\tau$ if it plays $\mathcal{NS}$. In other words, $APP$ is not indifferent between playing $\mathcal{S}$ and $\mathcal{NS}$ when $p \geq \frac{(1-p)\kappa}{\beta + \phi}$, and strongly prefers playing $\mathcal{S}$. Thus, there is no hybrid equilibria at $(\mathcal{NS},(\mathcal{S},\mathcal{NS}))$ when $p \geq \frac{(1-p)\kappa}{\beta + \phi}$.

The second possibility, in terms of the $DM$'s best response to an $APP$'s strategy of $\mathcal{NS}$, is for the $DM$ to Accept or $A$ (i.e., $A$ dominates $B$) which is true if $p \leq \frac{(1-p)\kappa}{\beta + \phi}$. This combined with the fact that the $DM$ always plays $A$ for $\mathcal{S}$, it is clear that when $p \leq \frac{(1-p)\kappa}{\beta + \phi}$, $DM$ invariantly plays $A$ for both the $\mathcal{S}$ and $\mathcal{NS}$ strategies of the $APP$. In this case, if $MA$ deviates and plays $\mathcal{S}$ it will gain $\alpha + u - c^\mathcal{S}$ compared to $\alpha - c^\mathcal{NS}$ if it plays $\mathcal{NS}$.
Similarly, $HA$ will gain $\sigma + v - c^\mathcal{S}$ instead of $\sigma - c^\mathcal{NS}$ if it plays $\mathcal{NS}$. Therefore, in order to make $APP$ indifferent between playing $\mathcal{S}$ and $\mathcal{NS}$ so that a hybrid equilibrium can be achieved at $(\mathcal{NS},(\mathcal{S},\mathcal{NS}))$, the following conditions must be satisfied:
{\scriptsize
$$c^\mathcal{S} - c^\mathcal{NS} \simeq u$$
$$c^\mathcal{S} - c^\mathcal{NS} \simeq v$$
}
In summary, a hybrid equilibrium is possible at $(\mathcal{NS},(\mathcal{S},\mathcal{NS}))$ if and only if the above conditions hold.
\end{proof}

\begin{theorem}\label{thm:hybrid-s-ns-s}
There exists a hybrid equilibrium on the $APP$ strategy profile $((\mathcal{S},\mathcal{NS}),\mathcal{S})$ in game $\mathbb{G}_D$.
\end{theorem}

\begin{proof}
An $APP$ strategy profile $((\mathcal{S},\mathcal{NS}),\mathcal{S})$ means that $HA$ always selects $\mathcal{S}$ (i.e., $n=1$), whereas $MA$ selects $\mathcal{S}$ with some probability $m$ and $\mathcal{NS}$ with probability $1-m$ where ($0<m<1$). $DM$'s beliefs in this strategy profile can thus be calculated as:
{\scriptsize
$$Pr(MA|\mathcal{S}) = q = \frac{Pr(\mathcal{S}|MA) \times Pr(MA)}{Pr(\mathcal{S})} = \frac{m \times \theta}{m \times \theta + 1 \times (1 - \theta)} = \frac{\theta m}{\theta(m-1) + 1}$$
$$Pr(MA|\mathcal{NS}) = p = \frac{Pr(\mathcal{NS}|MA) \times Pr(MA)}{Pr(\mathcal{NS})} = \frac{\theta \times (1-m)}{\theta \times (1-m) + 0 \times (1 - \theta)} = 1$$
}
Now, let's compute the $DM$'s best response for each of the strategies $\mathcal{S}$ and $\mathcal{NS}$ of $APP$. In order to determine that, we need to first compute the expected utilities/payoffs obtained by the $DM$ for playing $B$ or $A$ if $APP$ ($MA$ or $HA$) selects $\mathcal{NS}$ or $\mathcal{S}$, which is given by:
{\scriptsize
$$EU_{DM}(B,\mathcal{NS}) = 1 \times (\beta -\psi^\mathcal{NS}) + 0 \times (-\kappa -\psi^\mathcal{NS}) = \beta -\psi^\mathcal{NS}$$
$$EU_{DM}(A,\mathcal{NS}) = 1 \times (-\phi -\psi^\mathcal{NS}) +  0 \times (-\psi^\mathcal{NS}) = -\phi -\psi^\mathcal{NS}$$
$$EU_{DM}(B,\mathcal{S}) = q \times (\beta -\psi^\mathcal{S}) + (1-q) \times (-\kappa -\psi^\mathcal{S})$$
$$EU_{DM}(A,\mathcal{S}) = q \times (-\phi -\psi^\mathcal{S}) +  (1-q) \times (-\psi^\mathcal{S})$$
}
It is clear from these expected utilities obtained by the $DM$ in this strategy profile that it will always plays $B$ (i.e., $B$ always dominates $A$) when the $APP$ plays $\mathcal{NS}$. On the contrary, there are two possibilities in terms of the $DM$'s best response to an $APP$'s strategy of $\mathcal{S}$. The first possibility is for the $DM$ to always Block or $B$, i.e., $B$ would dominate $A$. This, however, holds only if the following is true: 
{\scriptsize
$$ q(\beta -\psi^\mathcal{S}) + (1-q)(-\kappa -\psi^\mathcal{S}) \geq  q(-\phi -\psi^\mathcal{S}) +  (1-q)(-\psi^\mathcal{S}) \equiv q \geq \frac{(1-q)\kappa}{\beta + \phi}$$
}
Now, as $DM$ always plays $B$ for $\mathcal{NS}$, $HA$ has no incentive to play $\mathcal{NS}$ because it will gain $-\gamma$ which is the same as what it would get if it plays $\mathcal{S}$. Similarly, $MA$ also has no incentive to play $\mathcal{NS}$ since it will gain $-\tau$ which is the same as what it would get if it plays $\mathcal{S}$. In other words, $APP$ is indifferent between playing $\mathcal{S}$ and $\mathcal{NS}$ when $q \geq \frac{(1-q)\kappa}{\beta + \phi}$. Thus, there is a hybrid equilibria at $((\mathcal{S},\mathcal{NS}),\mathcal{S})$ when $q \geq \frac{(1-q)\kappa}{\beta + \phi}$.

The second possibility, in terms of the $DM$'s best response to an $APP$'s strategy of $\mathcal{S}$, is for the $DM$ to Accept or $A$ (i.e., $A$ dominates $B$) which is true if $q \leq \frac{(1-q)\kappa}{\beta + \phi}$. This combined with the fact that the $DM$ always plays $B$ for $\mathcal{NS}$, it is clear that when $q \leq \frac{(1-q)\kappa}{\beta + \phi}$, both $MA$ and $HA$ will always play $\mathcal{S}$ as the payoff for playing $\mathcal{S}$ is always greater than switching. In other words, $APP$ is not indifferent between playing $\mathcal{S}$ and $\mathcal{NS}$ when $q \leq \frac{(1-q)\kappa}{\beta + \phi}$, and strongly prefers playing $\mathcal{S}$. Thus, there is no hybrid equilibria at $((\mathcal{S},\mathcal{NS}),\mathcal{S})$ when $q \leq \frac{(1-q)\kappa}{\beta + \phi}$.

In summary, a hybrid equilibrium is possible at $((\mathcal{S},\mathcal{NS}),\mathcal{S})$ if and only if $q \geq \frac{(1-q)\kappa}{\beta + \phi}$.
\end{proof}

\begin{theorem}\label{thm:hybrid-s-ns-ns}
There exists a hybrid equilibrium on the $APP$ strategy profile $((\mathcal{S},\mathcal{NS}),\mathcal{NS})$ in game $\mathbb{G}_D$.
\end{theorem}

\begin{proof}
An $APP$ strategy profile $((\mathcal{S},\mathcal{NS}),\mathcal{NS})$ means that $HA$ always selects $\mathcal{NS}$ (i.e., $n=0$), whereas $MA$ selects $\mathcal{S}$ with some probability $m$ and $\mathcal{NS}$ with probability $1-m$ where ($0<m<1$). $DM$'s beliefs in this strategy profile can thus be calculated as follows:
{\scriptsize
$$Pr(MA|\mathcal{NS}) = p = \frac{Pr(\mathcal{NS}|MA) \times Pr(MA)}{Pr(\mathcal{NS})} = \frac{(1 - m) \times \theta}{(1 - m) \times \theta + 1  \times (1 - \theta)} = \frac{\theta(1-m)}{\theta(1-m) + (1-\theta)}$$
$$Pr(MA|\mathcal{S}) = q = \frac{Pr(\mathcal{S}|MA) \times Pr(MA)}{Pr(\mathcal{S})} =  \frac{m \times \theta}{m \times \theta + 0 \times (1 - \theta)} = 1$$
}
Now, let's compute the $DM$'s best response for each of the strategies $\mathcal{S}$ and $\mathcal{NS}$ of $APP$. In order to determine that, we need to first compute the expected utilities/payoffs obtained by the $DM$ for playing $B$ or $A$ if $APP$ ($MA$ or $HA$) selects $\mathcal{S}$ or $\mathcal{NS}$, which is given by:
{\scriptsize
$$EU_{DM}(B,\mathcal{S}) = 1 \times (\beta -\psi^\mathcal{S}) + 0 \times (-\kappa -\psi^\mathcal{S}) = \beta -\psi^\mathcal{S}$$
$$EU_{DM}(A,\mathcal{S}) = 1 \times (-\phi -\psi^\mathcal{S}) +  0 \times (-\psi^\mathcal{S}) = -\phi -\psi^\mathcal{S}$$
$$EU_{DM}(B,\mathcal{NS}) = p \times (\beta -\psi^\mathcal{NS}) + (1-p) \times (-\kappa -\psi^\mathcal{NS})$$
$$EU_{DM}(A,\mathcal{NS}) = p \times (-\phi -\psi^\mathcal{NS}) +  (1-p) \times (-\psi^\mathcal{NS})$$
}
It is clear from these expected utilities obtained by the $DM$ in this strategy profile that it will always plays $B$ (i.e., $B$ always dominates $A$) when the $APP$ plays $\mathcal{S}$. On the contrary, there are two possibilities in terms of the $DM$'s best response to an $APP$'s strategy of $\mathcal{NS}$. The first possibility is for the $DM$ to always Block or $B$, i.e., $B$ would dominate $A$. This, however, holds only if the following is true: 
{\scriptsize
$$p (\beta -\psi^\mathcal{NS}) + (1-p)(-\kappa -\psi^\mathcal{NS}) \geq  p(-\phi -\psi^\mathcal{NS}) +  (1-p) (-\psi^\mathcal{NS})$$
$$\equiv p \geq \frac{(1-p)\kappa}{\beta + \phi}$$
}
Now, as $DM$ always plays $B$ for $\mathcal{S}$, $HA$ has no incentive to play $\mathcal{NS}$ because it will gain $-\gamma$ which is the same as what it would get if it plays $\mathcal{S}$. Similarly, $MA$ also has no incentive to play $\mathcal{NS}$ since it will gain $-\tau$ which is the same as what it would get if it plays $\mathcal{S}$. In other words, $APP$ is indifferent between playing $\mathcal{S}$ and $\mathcal{NS}$ when $p \geq \frac{(1-p)\kappa}{\beta + \phi}$. Thus, there is a hybrid equilibria at $((\mathcal{S},\mathcal{NS}),\mathcal{NS})$ when $p \geq \frac{(1-p)\kappa}{\beta + \phi}$.

The second possibility, in terms of the $DM$'s best response to an $APP$'s strategy of $\mathcal{NS}$, is for the $DM$ to Accept or $A$ (i.e., $A$ dominates $B$) which is true if $p \leq \frac{(1-p)\kappa}{\beta + \phi}$. This combined with the fact that the $DM$ always plays $B$ for $\mathcal{S}$, it is clear that when $p \leq \frac{(1-p)\kappa}{\beta + \phi}$, both $MA$ and $HA$ will always play $\mathcal{NS}$ as the payoff for playing $\mathcal{NS}$ is always greater than switching. In other words, $APP$ is not indifferent between playing $\mathcal{S}$ and $\mathcal{NS}$ when $p \leq \frac{(1-p)\kappa}{\beta + \phi}$, and strongly prefers playing $\mathcal{NS}$. Thus, there is no hybrid equilibria at $((\mathcal{S},\mathcal{NS}),\mathcal{NS})$ when $p \leq \frac{(1-p)\kappa}{\beta + \phi}$.

In summary, a hybrid equilibrium is possible at $((\mathcal{S},\mathcal{NS}),\mathcal{NS})$ if and only if $p \geq \frac{(1-p)\kappa}{\beta + \phi}$.
\end{proof}

\begin{theorem}\label{thm:mixed-static}
There exists a mixed strategy PBNE  in the game $\mathbb{G}_D$.
\end{theorem}

\begin{proof}
First, let's determine the conditions for each $APP$ type to randomize (or be indifferent) between its choices. Let's assume $DM$ plays the mixed strategy $(yB, (1 - y)A)$ for $\mathcal{S}$ (i.e., suspicious requests) and $(xB, (1 - x)A)$ for $\mathcal{NS}$ (i.e, non-suspicious requests). 
Then for the $APP$ type $MA$, the expected utilities/payoffs of playing $\mathcal{S}$ and $\mathcal{NS}$ are:   
{\scriptsize
$$EU_{MA}(\mathcal{S}) = y \times -\tau + (1-y) \times (\alpha + u - c^{S})$$
$$EU_{MA}(\mathcal{NS}) = x \times -\tau + (1-x) \times (\alpha - c^{NS})$$
}
$MA$ is indifferent between playing $\mathcal{S}$ and $\mathcal{NS}$ if $EU_{MA}(\mathcal{S})$ = $ EU_{MA}(\mathcal{NS})$,  which gives: 
{\scriptsize
\begin{eqnarray}
y(\tau + \alpha + u - c^{S}) - x(\tau + \alpha - c^{NS}) & = & u - c^{S} + c^{NS}
\label{eq:euMAequate}
\end{eqnarray}
}
Similarly, for the $APP$ type $HA$, the expected utilities/payoffs of playing $\mathcal{S}$ and $\mathcal{NS}$ are:
{\scriptsize
$$EU_{HA}(\mathcal{S}) = y \times -\gamma + (1-y) \times (\sigma + v - c^{S})$$
$$EU_{HA}(\mathcal{NS}) = x \times -\gamma + (1-x) \times (\sigma - c^{NS})$$
}
$HA$ is indifferent between playing $\mathcal{S}$ and $\mathcal{NS}$ if $EU_{HA}(\mathcal{S})$ = $ EU_{HA}(\mathcal{NS})$, which gives: 
{\scriptsize
\begin{eqnarray}
y(\gamma + \sigma + v - c^{S}) - x(\gamma + \sigma - c^{NS}) & = & v - c^{S} + c^{NS}
\label{eq:euHAequate}
\end{eqnarray}
}
Solving Equations \ref{eq:euMAequate} and \ref{eq:euHAequate} for $x$ and $y$, we get $DM$'s mixed strategy for which each $APP$ type is indifferent between playing $\mathcal{S}$ and $\mathcal{NS}$. Let this $x = x^{*}$ and $y = y^{*}$.

Now let's determine the conditions for $DM$ to randomize (or be indifferent) between its choices. First, if $DM$ observes $APP$ ($MA$ or $HA$) played $\mathcal{S}$, its expected payoffs from playing $B$ and $A$ are:
{\scriptsize
$$EU_{DM}(\mathcal{B}) = q \times (\beta - \psi^{S}) + (1-q) \times (-\kappa - \psi^{S})$$
$$EU_{DM}(\mathcal{A}) = q \times (-\phi - \psi^{S}) + (1-q) \times -\psi^{S}$$
}
Now, $DM$ is indifferent between playing $B$ and $A$ on seeing $\mathcal{S}$ if, $EU_{DM}(\mathcal{B})$ = $ EU_{DM}(\mathcal{A})$, which gives:
{\scriptsize
$$q = \frac{\kappa}{\kappa + \beta + \phi}\nonumber = q^{*}$$
}
Similarly, $DM$'s expected utilities/payoffs from playing $B$ and $A$, when it sees $\mathcal{NS}$ are: 
{\scriptsize
$$EU_{DM}(\mathcal{B}) = p \times (\beta - \psi^{NS}) + (1-p) \times (-\kappa - \psi^{NS})$$
$$EU_{DM}(\mathcal{A}) = p \times (-\phi - \psi^{NS}) + (1-p) \times -\psi^{NS}$$
}
$DM$ is indifferent between playing $B$ and $A$ on seeing $\mathcal{NS}$ if $EU_{DM}(\mathcal{B})$ = $ EU_{DM}(\mathcal{A})$, which gives:
{\scriptsize
$$p = \frac{\kappa}{\kappa + \beta + \phi}\nonumber = p^{*}$$
}
Now, we determine $APP$ ($MA$ or $HA$) randomization (mixed strategy) that is consistent with $DM$'s beliefs. 
For that, we use Bayes rule to calculate the $DM$'s beliefs $q$ and $p$ as: 
{\scriptsize
\begin{eqnarray}
q = q^{*} = \frac{m \times \theta}{m \times \theta + n \times (1-\theta)} \label{eq:qDMbelief}\\
p = p^{*} = \frac{(1 - m) \times \theta}{(1 - m) \times \theta + (1 - n) \times (1-\theta)} \label{eq:pDMbelief}
\end{eqnarray}
}
We can solve Equations \ref{eq:qDMbelief} and \ref{eq:pDMbelief} for $m$ and $n$, to obtain $MA$'s and $HA$'s mixed strategy for which they are indifferent in playing $\mathcal{S}$ and $\mathcal{NS}$ consistent with the $DM$'s beliefs. It is easy to show that there exists a system of (cost/benefit) parameters for which such a solution exists. Let these solutions be represented as $m^{*}$ and $n^{*}$. 
Then, the mixed strategy PBNE $\mu^{*}$ will occur at:\\ 
$\mu^{*}_{APP}$: $MA$ plays $(m^{*}\mathcal{S} + (1 - m^{*})\mathcal{NS})$ and HA plays $(n^{*}\mathcal{S} + (1 - n^{*})\mathcal{NS})$\\
$\mu^{*}_{DM}$: $DM$ plays $y^{*}B + (1 - y^{*})A$ to $\mathcal{S}$ and $x^{*}B + (1 - x^{*})A$ to $\mathcal{NS}$\\
$DM$'s beliefs: $q$ = Pr(MA|$\mathcal{S}$) =  $q^{*}$ and $p$  = Pr(MA|$\mathcal{NS}$) =  $p^{*}$

Example of a mixed equilibrium: Substituting $\theta=\frac{1}{2}$, $q=\frac{1}{4}$ and $p=\frac{3}{4}$ in Equations \ref{eq:qDMbelief} and \ref{eq:pDMbelief}, and solving for $m$ and $n$, results in $m=\frac{1}{4}$ and $n=\frac{3}{4}$.
\end{proof}

{
\def\arraystretch{1.50}
\begin{table}[h]
\scriptsize
\caption{List of  PBNEs.}
\begin{center}
\begin{tabular}{ |m{4cm}|m{2cm}|m{5.5cm}|}
\hline
{ \textbf{Conditions}}  & { \textbf{Range of $\theta$}} & { \textbf{PBNE Profiles}}   \\  \rowcolor[gray]{.9} \hline
$--$ &$\theta \geq \frac{\kappa}{\beta + \kappa + \phi}$ &$\mathcal{PBNE} = \{(\mathcal{S},\mathcal{S}), (B, B),p, q\}$ \\ \hline
$v \geq c^\mathcal{S} - c^\mathcal{NS}$, $u \geq c^\mathcal{S} - c^\mathcal{NS}$ &$\theta \leq \frac{\kappa}{\beta + \kappa + \phi}$ &$\mathcal{PBNE} = \{(\mathcal{S},\mathcal{S}), (A, A),p, q\}$ \\ \rowcolor[gray]{.9} \hline
$--$ &$\theta \leq \frac{\kappa}{\beta + \kappa + \phi}$ &$\mathcal{PBNE} = \{(\mathcal{S},\mathcal{S}), (A, B),p, q\}$ \\ \hline
$--$ &$\theta \geq \frac{\kappa}{\beta + \kappa + \phi}$ &$\mathcal{PBNE} = \{(\mathcal{NS},\mathcal{NS}), (B, B),p, q\}$ \\ \rowcolor[gray]{.9} \hline
$v \leq c^\mathcal{S} -  c^\mathcal{NS}$, $u \leq c^\mathcal{S} -  c^\mathcal{NS}$ &$\theta \leq \frac{\kappa}{\beta + \kappa + \phi}$ &$\mathcal{PBNE} = \{(\mathcal{NS},\mathcal{NS}), (A, A),p, q\}$ \\ \hline
$--$ &$\theta \leq \frac{\kappa}{\beta + \kappa + \phi}$ &$\mathcal{PBNE} = \{(\mathcal{NS},\mathcal{NS}), (B, A),p, q\}$ \\ \rowcolor[gray]{.9} \hline
$ c^\mathcal{S} - c^\mathcal{NS} \simeq u$, $c^\mathcal{S} - c^\mathcal{NS} \simeq v$ & $q \leq \frac{(1-q)\kappa}{\beta + \phi}$ &$\mathcal{PBNE} = \{(\mathcal{S},(\mathcal{S},\mathcal{NS})), (A, A), p, q\}$ \\ \hline
$ c^\mathcal{S} - c^\mathcal{NS} \simeq u$, $c^\mathcal{S} - c^\mathcal{NS} \simeq v$ &$p \leq \frac{(1-p)\kappa}{\beta + \phi}$ &$\mathcal{PBNE} = \{(\mathcal{NS},(\mathcal{S},\mathcal{NS})), (A, A), p, q\}$ \\ \rowcolor[gray]{.9} \hline
$--$ &$q \geq \frac{(1-q)\kappa}{\beta + \phi}$ &$\mathcal{PBNE} = \{((\mathcal{S},\mathcal{NS}),\mathcal{S}), (B, B), p, q\}$ \\ \hline
$--$ &$p \geq \frac{(1-p)\kappa}{\beta + \phi}$ &$\mathcal{PBNE} = \{((\mathcal{S},\mathcal{NS}),\mathcal{NS}), (B, B), p, q\}$ \\ \rowcolor[gray]{.9} \hline
\end{tabular}
\end{center}
\label{tab:Proofs}
\end{table}
}

This concludes our discussion of the different PBNEs in game $\mathbb{G}_D$ (summarized in Table \ref{tab:Proofs}).


\section{Numerical Analysis}
\label{sec:numericalanalysis}

We perform numerical simulations to analyze how the various PBNEs in our Sensor Access Signaling Game $\mathbb{G}_D$ evolves with respect to the various game and system parameters. 
Specifically, we evaluate the $MA$'s payoff, $HA$'s payoff and $DM$'s expected utility ($EU_{DM}$) in a representative separating strategy profile $(\mathcal{S},\mathcal{NS})$, a pooling strategy profile $(\mathcal{S},\mathcal{S})$, a hybrid strategy profile $((\mathcal{S},\mathcal{NS}),\mathcal{S})$ and a mixed strategy profile, by varying the value of $\theta$ (Nature's selection probability). The results are outlined in Figure \ref{numerical-graphs}, and the set of system parameters chosen for the numerical simulations are summarized in Figure \ref{para_table}. 

{
\begin{figure}
\centering
\subfloat[Separating Strategy $(\mathcal{S},\mathcal{NS})$\label{s_ns_theta-graph}]{
\includegraphics[width=.48\textwidth]{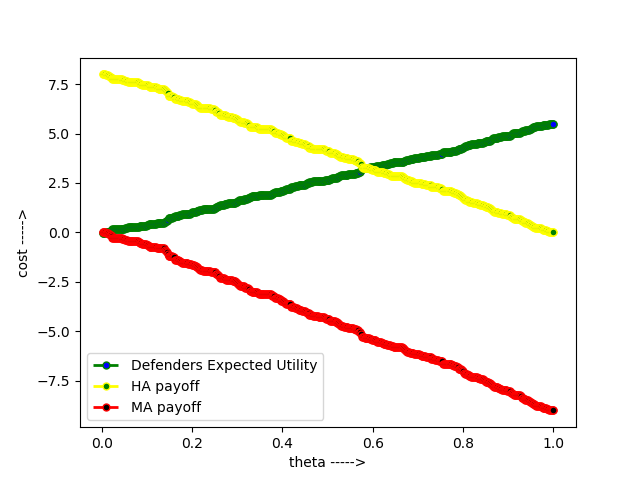}
}
\subfloat[Pooling Strategy $(\mathcal{S},\mathcal{S})$\label{s_s_theta-graph}]{
\includegraphics[width=.48\textwidth]{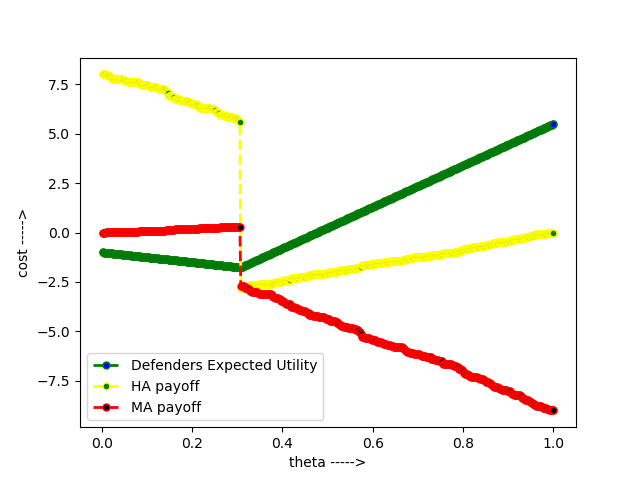}
}

\subfloat[Hybrid Strategy $((\mathcal{S},\mathcal{NS}),\mathcal{S})$\label{hybid_DM-graph}]{
\includegraphics[width=.48\textwidth]{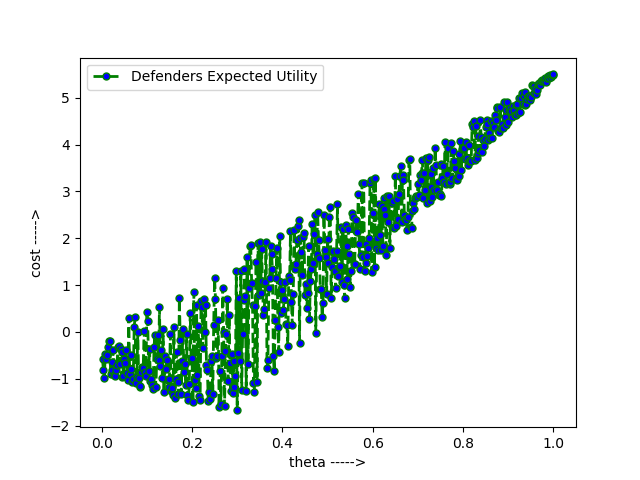}
}
\subfloat[Hybrid Strategy $((\mathcal{S},\mathcal{NS}),\mathcal{S})$\label{hybrid_MA-HA-graph}]{
\includegraphics[width=.48\textwidth]{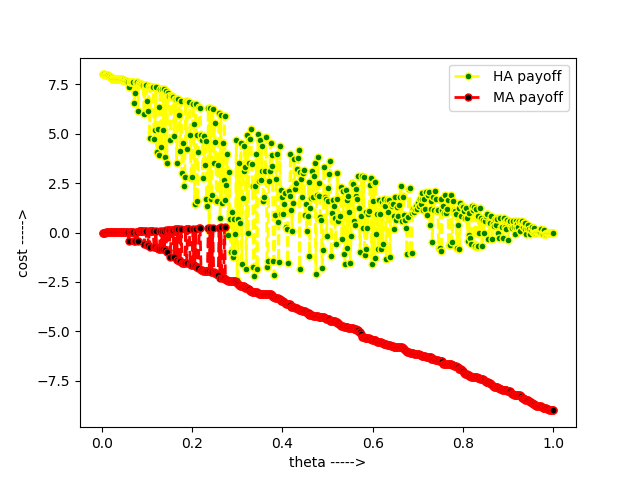}
}

\subfloat[Mixed Strategy\label{mixed}]{
\includegraphics[width=.48\textwidth]{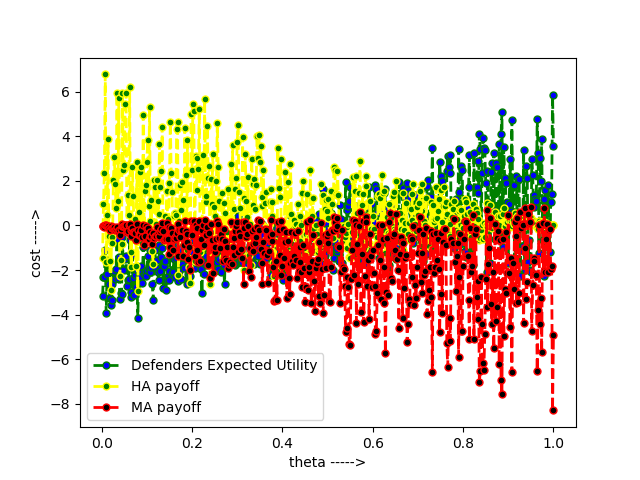}
}
\subfloat[Simulation Parameters\label{para_table}]{
\includegraphics[width=.48\textwidth]{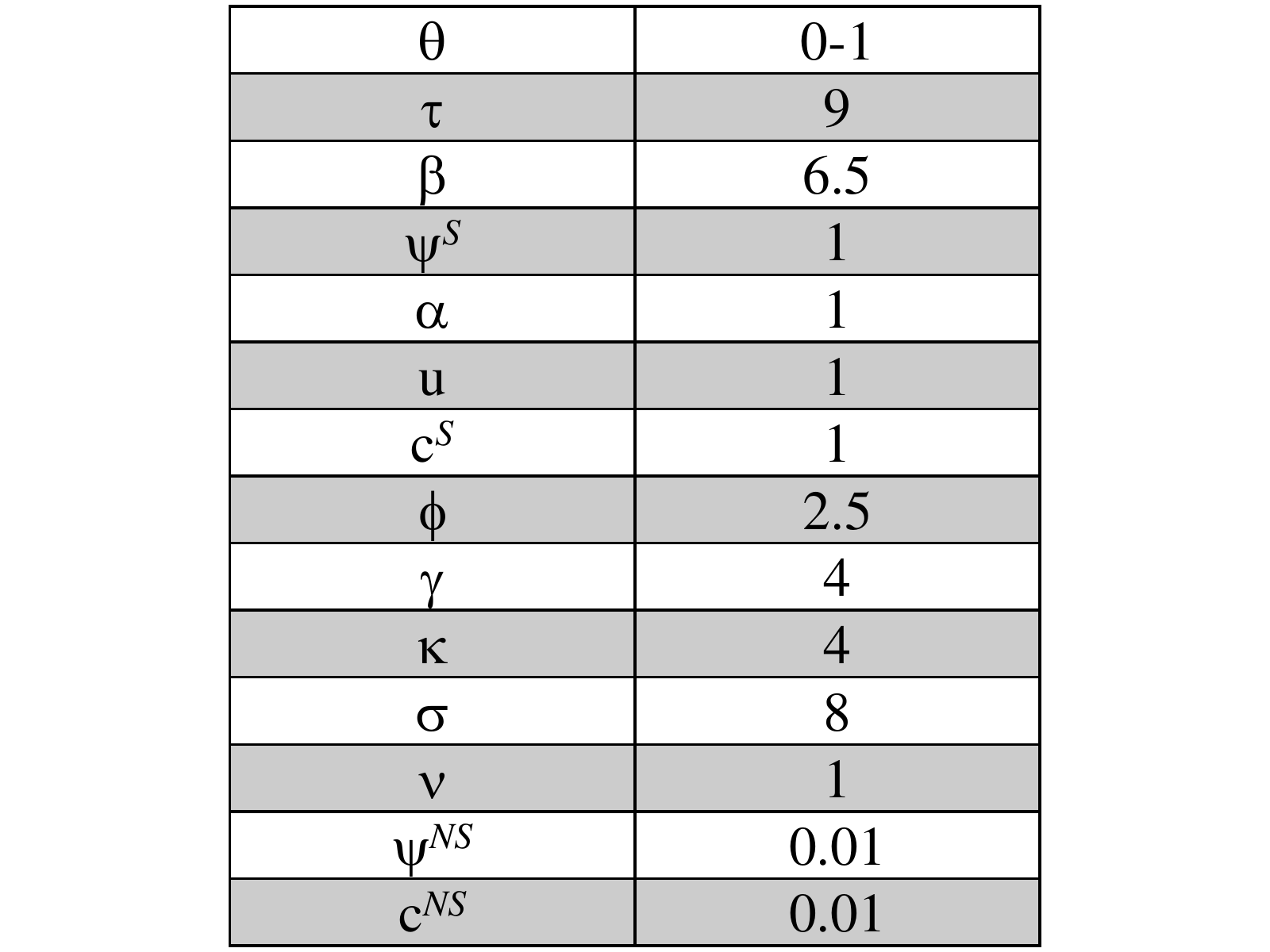}
}
\caption{(a-e) Effect of $\theta$ on different strategy profiles. Each point is a average of 500 iterations. (f) Default simulation parameters.}
\label{numerical-graphs}
\end{figure}
}



\noindent
\textbf{Separating strategy $(\mathcal{S},\mathcal{NS})$.} As proved earlier, there is no equilibrium in any of the separating strategy profiles, and the same can also be observed in the Figure \ref{s_ns_theta-graph}. We observe that $EU_{DM}$ is linearly increasing, which implies that $DM$ is blocking suspicious requests from $MA$, as the only way $DM$ can increase it's utility is by playing $B$. Both $MA$'s and $HA$'s payoffs are linearly decreasing because $DM$ is playing $B$ more than $A$.

\noindent
\textbf{Pooling strategy $(\mathcal{S},\mathcal{S})$. } In Figure \ref{s_s_theta-graph} we observe that the $HA$'s payoff and $DM$'s expected utility initially decreases while $MA$'s payoff increases, for increasing values of $\theta$. However, beyond a certain value of $\theta$ the trend reverses, i.e, $HA$'s payoff and $DM$'s expected utility increases linearly while $MA$'s payoff decreases.

\noindent
\textbf{Hybrid strategy $((\mathcal{S},\mathcal{NS}),\mathcal{S})$. } In this strategy profile (Figure \ref{hybid_DM-graph}), $EU_{DM}$ is affected by random signals coming from $MA$. However, we can also observe that as $\theta$ increases $EU_{DM}$ gradually increases. $EU_{DM}$ also stabilized for higher values of $\theta$. 
On the other hand, $HA$'s and $MA$'s payoffs are decreasing as expected when increasing $\theta$ (Figure \ref{hybrid_MA-HA-graph}).

\noindent
\textbf{Mixed strategy.} In Figure \ref{mixed} we observe the effect of a mixed strategy in each player's payoff/utility. The payoffs and utilities are highly unstable as $m$, $n$, $x$ and $y$ are all drawn from a random distribution for the mixed strategy.

In summary, our numerical evaluations validate our game-theoretic results.

\section{Repeated Game}
\label{sec:repeated}
So far, we have outlined PBNE results and related numerical analysis for the Sensor Access Signaling Game $\mathbb{G}_D$ in the single stage (or single-shot) scenario. In practice, however, the game $\mathbb{G}_D$ will be \emph{repeated} several times (possibly, as long as the system is running). Thus, it is important to analyze how the game $\mathbb{G}_D$ will evolve in a repeated scenario. 

\subsection{Background}
Before proceeding ahead, let us provide some technical background on repeated games. There are two broad categories of repeated games: 

\noindent
\textbf{(i) Finite Repeated Games:} Here, a stage game is repeated for a \emph{finite} number of times. Repeated games could support strategy profiles (also known as $reward$ and $punishment$ strategies) that support deviation from stage game Nash Equilibria through cooperation. Players could cooperate and play a reward strategy (also referred to as a \emph{Subgame Perfect Equilibrium} (SPE)) that is not a Nash Equilibrium strategy, if the expected utility of every player is strictly greater than the expected utility from the Nash Equilibrium strategy \cite{osborne1994course}. Due to the lower expected utility, the Nash Equilibrium strategy becomes the $punishment$ strategy, which would be applied if any of the players deviate from the SPE.
However, if a finite repeated game consists of stage games that each have a unique Nash Equilibrium, then the repeated game also has a unique SPE of playing the stage game Nash Equilibrium in each stage. This can be explained by \emph{unravelling} from the last stage, where players must play the unique Nash Equilibrium. In the second-to-last stage, as players cannot condition the future (i.e., the last stage) outcomes, again they must play the unique Nash Equilibrium for optimal expected utility. This backward induction continues until the first stage of the game, implying that players must always play the Nash Equilibrium strategy to ensure overall optimal expected utility. 
This (players not cooperating on a reward strategy) is a limitation of finite repeated games with a unique Nash Equilibrium, that can be solved if the game is repeated infinitely.

\noindent  
\textbf{(ii) Infinite Repeated Games:} 
In a repeated game with an infinite (or unknown) number of stages, players can condition their present actions upon the unknown future. Without a known end stage, players will be more inclined to cooperate on a mutually beneficial reward strategy, rather than a static Nash Equilibrium as seen in a finite repeated game. 
The payoff/utility for a player $i$ in an infinite repeated game can be computed by discounting the expected utilities in future stages using a \emph{discount factor} $\delta$ ($0\leq\delta\leq1$) as:
{\scriptsize
\begin{eqnarray}
u_{i} = u_{i}^{1} +\delta u_{i}^{2} + \delta^{2}u_{i}^{2} + \ldots + \delta^{t-1}u_{i}^{t-1} + \ldots  = \sum_{t=1}^{\infty}\delta^{t-1}u_{i}^{t} \nonumber
\end{eqnarray}
}
And, the average (normalized) expected utility for player $i$ is $(1-\delta)\sum_{t=1}^{\infty}\delta^{t-1}u_{i}^{t}$. In an infinitely repeated game, players can effectively employ a \textit{reward-and-punishment} strategy, but to do so each player must maintain a \emph{history} of the past actions taken by all players.
Let $H_{t}$ denote the set of all possible histories ($h_{t}$) of length $t$ and let $H = \cup_{t=1}^{\infty}H_{t}$ be the set of all possible histories. A pure strategy ($\omega_{i}$) for player $i$ is a mapping $\omega_{i} : H \to \Omega_{i}$ that maps histories ($H$) into player actions ($\Omega_{i}$) of the stage game. 
In an infinitely repeated game $\mathbb{G}(t,\delta)$ of $n$ players, a strategy profile $\omega = (\omega_{1}, ..., \omega_{n})$ is a Subgame Perfect Equilibrium (SPE) if and only if there is no player \textit{i} and no single history $h_{t-1}$ for which player \textit{i} would gain by deviating from $\omega_{i}( h_{t-1})$. Next, let us analyze the Sensor Access Signaling Game $\mathbb{G}_D$ for the infinite repeated scenario. 

\subsection{Repeated $\mathbb{G}_D$ with History: A Case Study}
\label{sec:history}

Let us analyze one of the possible scenarios of an infinitely repeated game $\mathbb{G}_D$($t$), where we assume $\{(\mathcal{S},\mathcal{NS}),\mathcal{NS}, (B, A), q, p\}$ as the $reward$ strategy and $\{(\mathcal{S},\mathcal{NS}),\mathcal{NS}, (B, B), q, p\}$ as the $punishment$ strategy. In this scenario, $HA$ may start sending $\mathcal{S}$ at a later point in the game in order to increase its payoff from $\sigma - c^{NS}$ to $\sigma + v - c^{S}$. However, as each player maintains a history of action sets for every player, as soon as $HA$ deviates from the SPE, $DM$ will enforce the $punishment$ strategy profile, thus blocking all the incoming requests whether it is $\mathcal{S}$ or $\mathcal{NS}$. $MA$ is randomizing between $\mathcal{S}$ and $\mathcal{NS}$ according to the feasible reward strategy profile, so it does not matter to $DM$ if $MA$ deviates or not. It is not logical to assume that $DM$ will deviate as it is $DM$'s responsibility to keep check on the deviations of $APP$. Moreover, each stage in the game $\mathbb{G}_D(t)$ is a sequential game, where $DM$ reacts to $APP$'s signal in every stage of the game.

After each stage of the game, the set of actions of player $APP$ and the corresponding responses of player $DM$ will be known to all players. Players may change their strategy after a certain period or stage, based on the history information until that stage. Figure \ref{with_history_EUHA_MA} shows the effect of history on the repeated games. We observe that $HA$'s utility fluctuates whenever it deviates from the cooperative reward strategy. With a strategy reset interval of 100 stages, we observe that $HA$'s utility follows a up-down pattern in every interval, reflective of a start with reward strategy, then $HA$'s deviation from reward strategy, and followed by $DM$'s switch to the punishment strategy. Overall, $MA$'s cumulative payoff is lower than $HA$'s cumulative payoff, which is desired in our system as we want the $DM$ to thwart $MA$ while allowing $HA$ to function normally.

We also study the effect of \emph{discount factor} $\delta$ (on the game $\mathbb{G}_D(t,\delta)$), which determines players' patience. If the value of $\delta$ is high, then there is a high chance that game is going to progress to the next stage, prompting player to cooperate on the reward strategy for longer. In Figure \ref{with_historydelta_EUDM_HA_MA}, we initially observe $HA$'s utility increasing and $MA$'s utility decreasing as per the reward strategy. However, as the game progresses, the cumulative utilities converge because (i) the utilities are heavily discounted, and (ii) players switch to the Nash Equilibrium strategy as a result of the discounted utility.

{
\begin{figure}[t]
\centering
\subfloat[Utilities with history.\label{with_history_EUHA_MA}]{
\includegraphics[width=.48\textwidth]{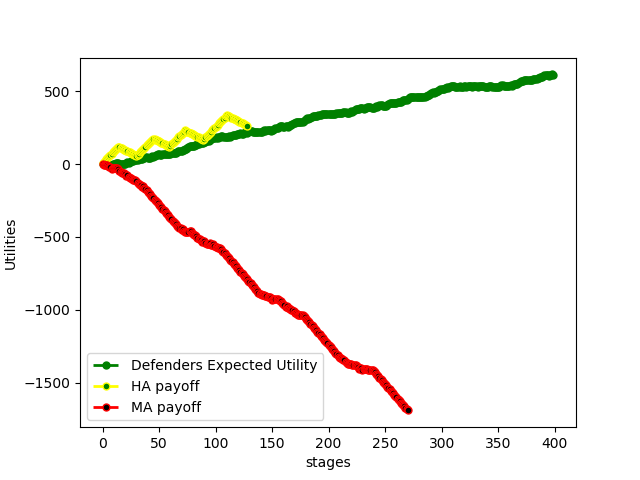}
}
\subfloat[Utilities with history and discount factor.\label{with_historydelta_EUDM_HA_MA}]{
\includegraphics[width=.48\textwidth]{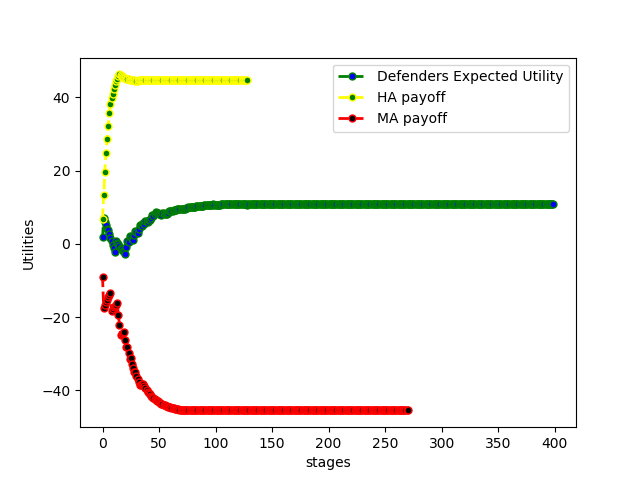}
}
\caption{Cumulative utilities for $DM$, $MA$, and $HA$ in repeated games.}
\label{repeated-graphs}
\end{figure}
}

\section{Related Work}
\label{sec:relatedwork} 

Several recent works demonstrated the feasibility of side-channel inference attacks using mobile 
\cite{FeltFCHW:2011,SchlegelZZIKW:2011,CaiC:2011,OwusuHDPZ:2012,MarquardtVCT:2012,NguyenCWBZ:2012,HanONPZ:2012,MiluzzoVBC:2012,GaoFSKYL:2014,MichalevskyDG:2014,michalevsky2015powerspy,narain2016inferring} and wearable \cite{Liu:2015:GBE:2810103.2813668,wang2015,maiti:2016,Wang:2016:FFY:2897845.2897847,MaitiJHB:2018,SabraMJ:2018,MaitiHSJ:2018} device sensors. Some of these works also propose defense mechanism against the specific type of attack that was demonstrated. For example, Miluzzo et al. \cite{MiluzzoVBC:2012} proposed to drastically reduce the maximum allowed sensor sampling rate, in order to prevent keystroke inference attacks on mobile keypads using mobile device motion sensors. However, reducing the sensor sampling rate for all applications may cause certain applications to malfunction, leading to poor user experience. To minimize unnecessary regulation of sensors at all times, Maiti et al. \cite{maiti:2016} proposed an activity recognition-based defense framework. In their framework, the defense mechanism continuously monitors user's current activity (using smartwatch motion sensors data), and regulates third party applications' access to motion sensor only when typing activity is detected (in order to prevent keystroke inference). However, while such ad-hoc defense approaches are effective in preventing a specific type of attack, they may not be effective against other types of side-channel attacks. In this work, we generalize the problem of side-channel attacks using mobile and wearable sensors, by modeling all different types of attacks as a Bayesian signaling game between a mobile application and a defense mechanism.

Bayesian signaling games to model malicious behavior has been used before in other research areas. For example, Patcha et al. \cite{patcha2004game} modeled a game for intrusion detection in mobile ad-hoc networks, however, they did not derive the equilibria of the game. Liu et al. \cite{liu2006bayesian} derived only the mixed-strategy Nash equilibria of a similar game of intrusion detection in mobile ad-hoc networks, using a belief updating scheme. A key difference between their game and ours is that in their game a ``regular'' player is assumed to be non-malicious at all times, which in other words mean that the game does not consider false positives. We did not include this assumption because an honest application's useful tasks may benefit from sending seeming suspicious sensor access requests, as captured by the variable $v$ in our game model.



\section{Conclusion}
\label{sec:conclusion} 


In this paper, we modeled the problem of zero-permission sensor access control for mobile applications using game theory. By means of a formal and practical signaling game model, we proved conditions under which equilibria can be achieved between entities with conflicting goals in this setting, i.e., honest and malicious applications who are requesting sensor access to maximize their utility and attack goals, respectively, and the defense mechanism who wants to protect against attacks without compromising system utility. By means of numerical simulations, we further studied how the different theoretically derived equilibria will evolve  in terms of the payoffs received by the application and the defense mechanism
Our results in this paper have helped shed light on how a defense mechanism can act in a strategically optimal manner to protect the mobile system against malicious applications that take advantage of zero-permission sensors to leak private user information and are impossible to detect otherwise.


\bibliographystyle{splncs04}
\bibliography{citations}

\end{document}